\documentclass[traditabstract]{aa} 

\usepackage{graphicx}
\usepackage{txfonts}
\usepackage{longtable}

\newcommand{\FOV}{field-of-view\ }
\newcommand{\los}{line-of-sight\ }
\begin{document}

   \title{{A Virgo Environmental Survey Tracing Ionised Gas Emission (VESTIGE).IV. A tail of Ionised Gas in the Merger Remnant NGC 4424\thanks{Based on observations obtained with
   MegaPrime/MegaCam, a joint project of CFHT and CEA/DAPNIA, at the Canada-French-Hawaii Telescope
   (CFHT) which is operated by the National Research Council (NRC) of Canada, the Institut National
   des Sciences de l'Univers of the Centre National de la Recherche Scientifique (CNRS) of France and
   the University of Hawaii. Based on observations obtained with XMM-Newton, an ESA science mission with instruments and contributions directly funded by ESA Member States and NASA.
   Based on observations made with ESO Telescopes at the La Silla Paranal Observatory under programmes ID 097.D-0408(A) and 098.D-0115(A). 
   Based  on  observations  collected  at  the  Observatoire  de  Haute Provence (OHP) (France), operated by the CNRS. Based on observations obtained at the Southern Astrophysical Research (SOAR) 
   telescope, which is a joint project of the Minist\'erio da Ci\^encia, Tecnologia, Inova\c{c}\~{o}es e Comunica\c{c}\~{o}es do Brasil (MCTIC/LNA), the U.S. National Optical Astronomy Observatory (NOAO), 
   the University of North Carolina at Chapel Hill (UNC), and Michigan State University (MSU).}}
   }
   \subtitle{}
  \author{A. Boselli\inst{1}\thanks{Visiting Astronomer at NRC Herzberg Astronomy and Astrophysics, 5071 west Saanich Road, Victoria, BC, V9E 2E7, Canada},         
          M. Fossati\inst{2,3},
          G. Consolandi\inst{4},
          P. Amram\inst{1},   
	  C. Ge\inst{5},
	  M. Sun\inst{5},
	  J.P. Anderson\inst{6},
          S. Boissier\inst{1},
	  M. Boquien\inst{7},
	  V. Buat\inst{1},
	  D. Burgarella\inst{1},               
          L. Cortese\inst{8,9},   
          P. C{\^o}t{\'e}\inst{10},
	  J.C. Cuillandre\inst{11},
          P. Durrell\inst{12},
	  B. Epinat\inst{1},
          L. Ferrarese\inst{10},
          M. Fumagalli\inst{3},
	  L. Galbany\inst{13},  
          G. Gavazzi\inst{4},
	  J.A. G\'omez-L\'opez\inst{1},
          S. Gwyn\inst{10},      
          G. Hensler\inst{14},
	  H. Kuncarayakti\inst{15,16},
	  M. Marcelin\inst{1},
	  C. Mendes de Oliveira\inst{17},
	  B.C. Quint\inst{18},
	  J. Roediger\inst{10},
	  Y. Roehlly\inst{19},
	  S.F. Sanchez\inst{20},
	  R. Sanchez-Janssen\inst{21},
          E. Toloba\inst{22,23},
	  G. Trinchieri\inst{24},
	  B. Vollmer\inst{25}
       }

\institute{     
                Aix Marseille Univ, CNRS, CNES, LAM, Marseille, France
             \email{alessandro.boselli@lam.fr}
        \and  
                Max-Planck-Institut f\"{u}r Extraterrestrische Physik, Giessenbachstrasse, 85748, Garching, Germany 
	\and
                Institute for Computational Cosmology and Centre for Extragalactic Astronomy, Department of Physics, Durham University, South Road, Durham DH1 3LE, UK
		\email{matteo.fossati@durham.ac.uk}             
       	\and
                Universit\'a di Milano-Bicocca, piazza della scienza 3, 20100, Milano, Italy
                \email{guido.consolandi@mib.infn.it} 
	\and
	        Department of Physics and Astronomy, University of Alabama in Huntsville, Huntsville, AL 35899, USA
        \and
		European southern Observatory, Alonso de C\'ordova 3107, Casilla 19, Santiago, Chile 
	\and
		Centro de Astronom\'a (CITEVA), Universidad de Antofagasta, Avenida Angamos 601, Antofagasta, Chile
	\and
                International Centre for Radio Astronomy Research, The University of western Australia, 35 Stirling Highway, Crawley WA 6009, Australia
	\and
		ARC Centre of Excellence for All Sky Astrophysics in 3 Dimensions (ASTRO 3D)
        \and
                NRC Herzberg Astronomy and Astrophysics, 5071 west Saanich Road, Victoria, BC, V9E 2E7, Canada
	\and
                AIM, CEA, CNRS, Universit\' Paris-Saclay, Universit\'e Paris Diderot, Sorbonne Paris Cit\'e, Observatoire de Paris, PSL University, F-91191 Gif-sur-Yvette Cedex, France
        \and
                Department of Physiscs and Astronomy, Youngstown State University, Youngstown, OH, USA
        \and
		PITT PACC, Department of Physics and Astronomy, University of Pittsburgh, Pittsburgh, PA 15260, USA
	\and
                Department of Astrophysics, University of Vienna, Turkenschanzstrasse 17, 1180, Vienna, Austria
	\and
		Finnish Centre for Astronomy with ESO (FINCA), FI-20014 University of Turku, Finland
	\and	
		Tuorla Observatory, Department of Physics and Astronomy, FI-20014 University of Turku, Finland
        \and
		Instituto de Astronomia, Geof\'isica e Ci\^encias Atmosf\'ericas da Universidadede S\~{a}o Paulo, Cidade Universit\'aria, CEP:05508-900, S\~{a}o Paulo, SP, Brazil
        \and
		Cerro Tololo Inter-American Observatory, SOAR Telescope, Casilla 603, La Serena, Chile
        \and
                Univ. Lyon1, ENS de Lyon, CNRS, Centre de Recherche Astrophysique de Lyon UMR5574, 69230, Saint-Genis-Laval, France
        \and
		Instituto de Astronom\'ia, Universidad Nacional Aut\'onoma de  M\'exico, Circuito Exterior, Ciudad Universitaria, Ciudad de M\'exico 04510,  Mexico
	\and
		UK Astronomy Technology Centre, Royal Observatory Edinburgh, Blackford Hill, Edinburgh EH9 3HJ, UK
	\and
                UCO/Lick Observatory, University of California, Santa Cruz, 1156 High Street, Santa Cruz, CA 95064, USA
        \and
                Texas Tech University, Physics Department, Box 41051, Lubbock, TX 79409-1051, USA
	\and
		INAF-Osservatorio Astronomico di Brera, via Brera 28, 20121 Milano, Italy 
	\and
		Observatoire Astronomique de Strasbourg, UMR 7750, 11, rue de l'Universit\'e, 67000, Strasbourg, France
                }
               
\authorrunning{Boselli et al.}
\titlerunning{A tail of Ionised Gas in the Merger Remnant NGC 4424}

   \date{}

 
  \abstract  
{We have observed the late-type peculiar galaxy NGC 4424 during the Virgo Environmental Survey Tracing Galaxy Evolution (VESTIGE), 
a blind narrow-band H$\alpha$+[NII] imaging survey of the Virgo cluster 
carried out with MegaCam at the Canada French Hawaii Telescope (CFHT). The presence of a $\sim$ 110 kpc long (in projected distance) HI tail
in the southern direction indicates that this galaxy is undergoing a ram pressure 
stripping event. The deep narrow-band image revealed the presence of a low surface brightness 
($\Sigma (H\alpha)$ $\simeq$ 4 $\times$ 10$^{-18}$ erg s$^{-1}$ cm$^{-2}$ arcsec$^{-2}$) ionised gas tail $\sim$ 10 kpc long 
extending from the centre of the galaxy to the north-west direction, thus in the direction opposite to the HI tail. 
\textit{Chandra} and \textit{XMM} X-rays data do not show any compact source in the nucleus nor the presence of an extended tail of hot gas,
while IFU spectroscopy (MUSE) indicates that the gas is photo-ionised in the inner regions
and shock-ionised in the outer parts.
Medium- (MUSE) and high-resolution (Fabry-Perot) IFU spectroscopy confirms that the ionised gas is kinematically decoupled from the stellar component 
and indicates the presence of two kinematically distinct structures in the stellar disc.
The analysis of the SED of the galaxy indicates that the activity of star formation has been totally quenched in the outer disc
$\sim$ 250-280 Myr ago, while only reduced by$\sim$ 80\% in the central regions. 
All this observational evidence suggests that NGC 4424 is the remnant of an unequal-mass merger occurred $\lesssim$ 500 Myr ago, 
when the galaxy was already a member of the Virgo cluster, now
undergoing a ram pressure stripping event which has removed the gas and quenched the activity of star formation in the outer disc.
The tail of ionised gas probably results from the outflow produced by a central starburst 
fed by the collapse of gas induced by the merging episode. This outflow is sufficiently powerful to overcome the ram pressure induced by 
the intracluster medium on the disc of the galaxy crossing the cluster. This analysis thus suggests that feedback can participate in the quenching process
of galaxies in high-density regions. 
 }
   {}
   {}
   {}
   {}
   {}


   \maketitle
%

\section{Introduction}

It is well established that the environment plays a major role in shaping galaxy
evolution. Rich clusters of galaxies in the local (Dressler 1980; Whitmore et al. 1993)
and intermediate redshift Universe (Dressler et al. 1997) are dominated by a quiescent galaxy population 
mainly composed of ellipticals and lenticulars. At the same time, star forming systems in rich environments 
are generally deprived of their interstellar medium (ISM) in its different phases, from the atomic (Haynes et al. 1984; Solanes et al. 2001; Vollmer et al. 2001a;
Gavazzi et al. 2005, 2006a, 2013) and molecular gas (Fumagalli et al. 2009; Boselli et al. 2014a) to the interstellar dust (Cortese et al. 2010, 2012a),
when compared to their field counterparts. As a consequence, their star formation could be significantly reduced 
(Kennicutt 1983; Gavazzi et al. 1998, 2002, 2006b, 2013; Boselli et al. 2014b, 2015).

Different physical processes have been proposed in the literature to explain the differences observed, as reviewed
in Boselli \& Gavazzi (2006, 2014). These processes can be broadly divided in two main families: those
related to the gravitational perturbation that a galaxy can suffer in a dense environment (galaxy-galaxy
interactions - Merritt 1983; galaxy-cluster interactions - Byrd \& Valtonen 1990; galaxy harassment - Moore
et al. 1998), and those due to the interaction of the galaxy ISM with the hot ($T$ $\simeq$ 10$^7$-10$^8$ K)
and dense ($\rho_{ICM}$ $\simeq$ 10$^{-3}$ cm$^{-3}$; Sarazin 1986) intra cluster medium (ICM) (ram pressure - Gunn \& Gott 1972; thermal evaporation - Cowie \& Songaila 1977; 
viscous stripping - Nulsen 1982; starvation - Larson et al. 1980).

The identification of the dominant mechanism responsible for the morphological transformation of galaxies in
dense environments is one of the major challenges of modern extragalactic astronomy. Indeed, it is expected
that the relative importance of the aforementioned processes varies with the mass of galaxies and of the
overdense regions, and thus with cosmic time given that these structures are growing with time. 
The identification of the principal process can be carried on by comparing complete samples of galaxies in
different environments, from voids to rich clusters,  
and at different redshifts, with the prediction of cosmological simulations and semi-analytic models of
galaxy evolution. It can also be accomplished by comparing the detailed multifrequency observations of 
representative objects to the predictions of models and simulations especially tailored to take into account
the effects of the different perturbations. This second approach has been succesfully
applied to a large number of galaxies in nearby clusters where the sensitivity and the angular resolution of
multifrequency observations allows us to compare in a coherent study the properties of the different galaxy
components (e.g. stellar populations, dust, gas in its different phases, magnetic fields) to the predictions
of models. Typical examples are the studies of several galaxies in the Virgo cluster (Kenney \& Koopmann 1999;
Kenney \& Yale 2002; Boselli et al. 2005, 2006; Kenney et
al. 2004, 2008, 2014; Vollmer et al. 1999, 2000, 2004a,b, 2005a,b, 2006, 2008a,b, 2009, 2012; 
Fumagalli et al. 2011; Abramson \& Kenney 2014; Abramson et al. 2011, 2016; Jachym et al. 2013), 
Coma cluster (Vollmer et al. 2001b; Kenney et al. 2015), A1367 (Gavazzi et al. 1995, 2001a,b; Consolandi et al. 2017), 
and Norma cluster (Sun et al. 2006, 2007, 2010; Fumagalli et al. 2014; Jachym et al. 2014;  Fossati et al. 2016),
or the jellyfish galaxies of the GASP survey (Poggianti et al. 2017). 
The analysis of these representative objects has been important not only to
understand the very nature of the physical mechanisms in act, but also to see that different processes can 
jointly participate and compete to the ongoing trasformation (Gavazzi et al. 2001a,b, 2003a; Vollmer 2003; Vollmer et al.
2005b; Cortese et al. 2006; Fritz et al. 2017).   

\begin{figure*}
   \centering
   \includegraphics[width=1\textwidth]{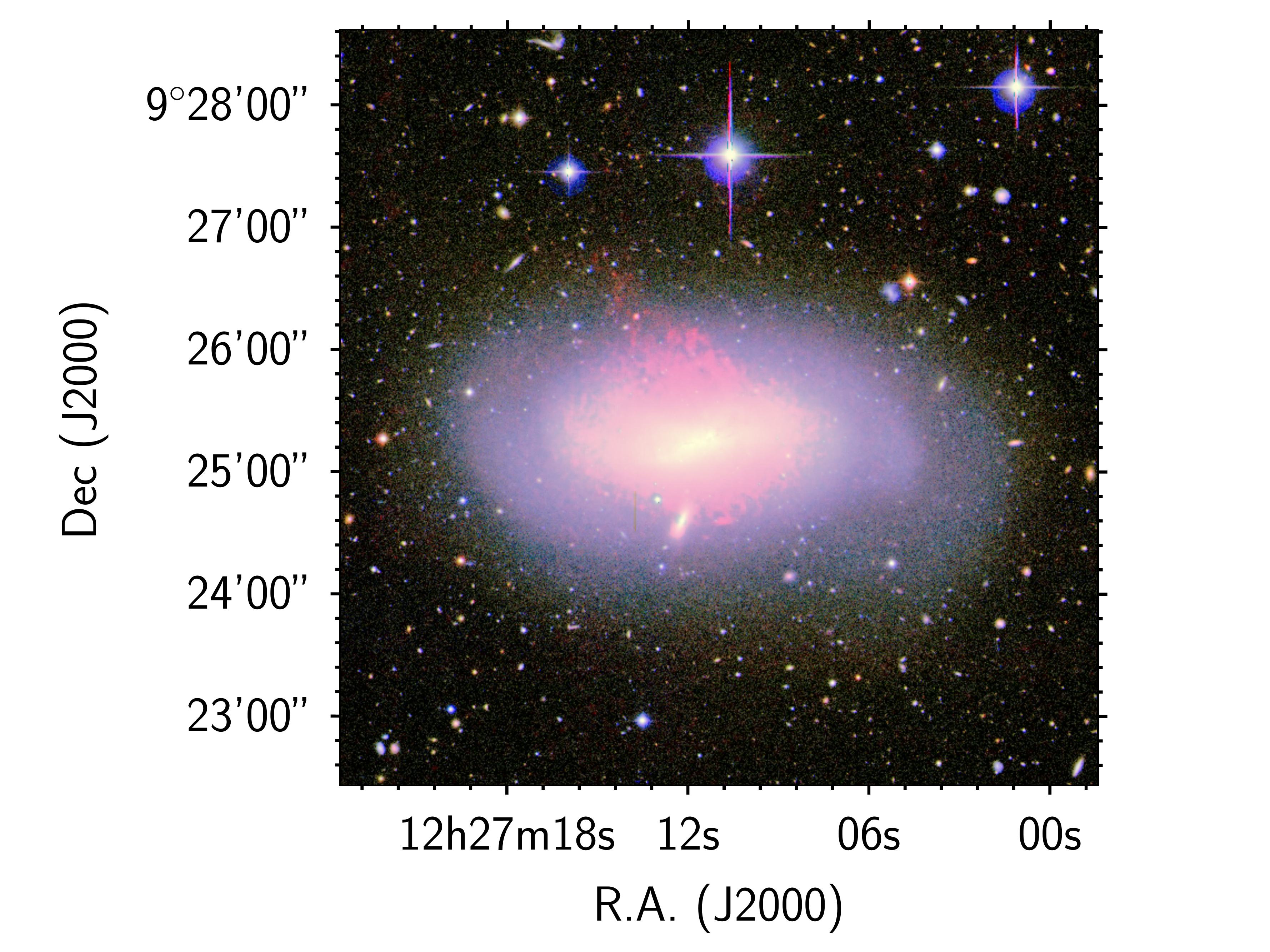}
   \caption{Pseudo-colour image of NGC 4424 obtained combining the NGVS (Ferrarese et al. 2012) optical $u$ and $g$ in the blue channel, the $r$ and NB in the green, and the $i$ and the 
   continuum-subtracted H$\alpha$ in the red. At the assumed distance of the galaxy (16.5 Mpc), 1 arcmin = 4.8 kpc. Stellar shells are not detected down to the 
   NGVS surface brightness limit of $\mu_g$ = 29 mag arcsec$^{-2}$.
 }
   \label{NGVS}%
   \end{figure*}

NGC 4424, the galaxy analysed in this work, is a typical example of an object undergoing a violent transformation
due to different mechanisms (Fig. \ref{NGVS} and Table \ref{gal}). 
This star forming system is located at $\simeq$ 0.9 Mpc ($\simeq$ 0.6 $R_{200}$) projected distance from M87 and  M49, the two dominant 
substructures of the Virgo cluster, but it is likely a member of the main cluster A given its distance of $\simeq$ 15.5-15.8 Mpc
(Munari et al. 2013, Hatt et al. 2018)\footnote{Consistently with other VESTIGE works, throughout this work we assume for NGC 4424
a distance of 16.5 Mpc as for all galaxies within the Virgo cluster A substructure (Gavazzi et al. 1999; Mei et al. 2007).}. The first analysis on optical data done by
Kenney et al. (1996) revealed a peculiar morphology with the presence of shells suggesting that this object has recently undergone 
a merging event. 
This scenario was later confirmed by high-resolution spectroscopic and CO observations by Cortes et al. (2006, 2015). 
The presence of a 110 kpc long tail in projected length of HI gas has been interpreted as a clear evidence that the
galaxy is now undergoing a ram pressure stripping event (Chung et al. 2007, 2009; Sorgho et al. 2017). Radio continuum observations also indicate
the presence of two polarised cones perpendicular to the disc of the galaxy probably due to a nuclear outflow (Vollmer et
al. 2013).

With the purpose of studying the effects of the environment on galaxy evolution, we are undertaking 
VESTIGE (A Virgo Environmental Survey Tracing Ionised Gas Emission), a deep
blind H$\alpha$+[NII]\footnote{Hereafter we will refer to the H$\alpha$+[NII] band simply as H$\alpha$,
unless otherwise stated.} narrow-band imaging survey of the Virgo cluster at the CFHT (Boselli et al. 2018a; paper I).
One of the purposes of this survey is identifying galaxies undergoing a perturbation with the hostile cluster environment 
through the observation of peculiar morphologies in the ionised gas component.
Indeed, deep blind observations of galaxies in nearby clusters, now made possible by the advent of large panoramic
detectors coupled with 4-8 metre class telescopes, have indicated H$\alpha$ narrow-band imaging as one of
the most efficient techniques for this purpose (Gavazzi et al. 2001a; Yoshida et al. 2002; Yagi et al. 2007, 2010, 2017; Sun et al. 2007; Kenney et al.
2008; Fossati et al. 2012; Zhang et al. 2013; Boselli et al. 2016a). The new H$\alpha$ image of NGC 4424 taken as part of the VESTIGE 
survey reveals the presence of spectacular features in the ionised gas distribution. We use these new data, combined with \textit{Chandra} and \textit{XMM} 
imaging and high-resolution 2D Fabry-Perot spectroscopic data gathered during follow-up observations and integral field spectroscopic data from VLT/MUSE  
to further analyse the kinematical and physical properties of this intriguing object. The observations and data reduction
are described in Sect. 2, while the description of the multifrequency data used in the analysis are in Sect. 3. The
derivation of the physical and kinematical parameters is given in Sect. 4, while the discussion and conclusions follow in Sect. 5 and 6.

\begin{table}
\caption{Properties of the galaxy NGC 4424 (VCC 979)}
\label{gal}
{
\[
\begin{tabular}{ccc}
\hline
\noalign{\smallskip}
\hline
Variable      		& Value			& Ref.          \\      
\hline
$Type$		& SB(s)a:HII						& 1	\\
$cz$		& 438 km s$^{-1}$					& 2	\\
$M_{star}^a$   	& 10$^{10.17}$ M$_{\odot}$				& 3	\\
$M(HI)$		& 10$^{8.35}$ M$_{\odot}$				& 2	\\
$M(H_2)$	& 10$^{8.57}$ M$_{\odot}$				& 2	\\
$Distance^b$	& 16.5 Mpc						& 4,5,6,7	\\
$incl$		& 62$^o$						& 8	\\
$Proj.~ distance~from~M87$	& 0.9 Mpc					& T.W.	\\
log$f(H\alpha+[NII])$      & -12.12$\pm$0.02  erg s$^{-1}$ cm$^{-2}$    & T.W.  \\
$SFR^a$		& 0.25	M$_{\odot}$yr$^{-1}$				& T.W.	\\
\noalign{\smallskip}
\hline
\end{tabular}
\]
References: 1) NED; 2) Boselli et al. 2014c; 3) Boselli et al. (2015); 4) Mei et al. (2007); 5) Gavazzi et al. (1999); 6) Blakeslee et al. (2009); 7) Cantiello et al. (2018); 7) Cortese et al. (2012b).\\
Notes: a) $M_{star}$ and $SFR$ are derived assuming a Chabrier (2003) IMF and the Kennicutt (1998) calibration; b) assumed distance for the determination of the physical parameters.
}
\end{table}

\section{Observations and data reduction}

\subsection{Narrow-band imaging}

The galaxy NGC 4424 has been observed during VESTIGE pilot observations in 
spring 2016. The observations have been carried out using MegaCam at the CFHT coupled with the new narrow-band (NB) filter MP9603 
($\lambda_c$ = 6590 \AA; $\Delta \lambda$ = 104 \AA) perfectly suited for the observations of Virgo cluster galaxies. At the redshift of the galaxy 
($cz$ = 438 km s$^{-1}$) the filter encompasses the H$\alpha$ line ($\lambda$ = 6563 \AA), where the transmissivity is $T$ = 92\%, 
and the two [NII] lines ($\lambda$ = 6548, 6583 \AA). MegaCam is composed of 40 CCDs with a pixel scale of 0.187 arcsec/pixel. 
As for the other pilot observations (Boselli et al. 2016a, 2018b; Fossati et al. 2018), the galaxy has been observed using 
the pointing-macro QSO LDP-CCD7 especially designed for the Elixir-LSB data reduction pipeline. 
This macro is composed of 7 different pointings overlapping over a region of 40$\times$30 arcmin$^2$. The determination of the stellar 
continuum is secured via a similar observation in the $r$-band filter. 
The integration time for each single pointing within the macro was of 660 sec in the NB filter and 66 sec in the $r$-band. 
The macro was run twice, thus the total integration time is of 924 sec in $r$-band and 9240 sec in NB. 
The typical sensitivity of the images is similar to that reached during the VESTIGE survey, i.e.
$f(H\alpha)$ $\sim$ 4$\times$10$^{-17}$ erg s$^{-1}$ cm$^{-2}$ for point soruces (5$\sigma$) and 
$\Sigma(H\alpha)$ $\sim$ 2$\times$10$^{-18}$ erg s$^{-1}$ cm$^{-2}$ arcsec$^{-2}$ for extended sources (1$\sigma$ after smoothing at 3 arcsec resolution; see
paper I for details). The images are of excellent quality, with a median image quality of 0.63 arcsec in the $r$-band and 0.62 arcsec in 
the NB filter. 

As for the other VESTIGE observations, the MegaCam images have been reduced using Elixir-LSB (Ferrarese et al. 2012), a data reduction pipeline especially
designed to detect the diffuse emission of extended low surface brightness features associated to perturbed cluster galaxies.
This pipeline efficiently removes any contribution of the scattered light in the different science frames 
as demonstrated by the analysis of broad-band images taken during the NGVS 
(Ferrarese et al. 2012; Mihos et al. 2015) survey. Pilot observations of the galaxy NGC 4569 (Boselli et al. 2016a) have shown that Elixir-LSB 
is perfectly suited for the NB frames whenever the images are background dominated, such as those under discussion.
The photometric calibration in both filters has been done following the procedures given in Fossati et al. (in prep.), with 
a typical uncertainty of 0.02-0.03 mag in both bands. A further check on the spectroscopic calibration of the images has 
been done by comparing the emission in the NB filter to that obtained by MUSE on the $\sim$ 1$\times$1 arcsec$^2$ gas emitting region.
This comparison gives consistent results within 3.5\%. The astrometry of each single image 
has been corrected using the MegaPipe pipeline (Gwyn 2008) before stacking. 

\subsection{X-rays imaging}

{\em Chandra} observation of NGC 4424 have been taken on April 17, 2017 with the Advanced CCD Imaging Spectrometer (ACIS). ACIS chip S3 is on the aimpoint (ObsID: 19408; PI: Soria). 
The image has no background flares and the clean time for the S3 chip is 14.6 ks. There were two {\em XMM} observations of NGC 4424. The first one (0651790101; PI: Sun) 
has been taken on June 13, 2010, but was completely ruined by background flares. The second one (0802580201; PI: Sun) has been gathered on December 5, 2017, with an integration time of 34 ks. 
Once removed the background flares, the clean time is 14.1 ks for MOS1, 13.8 ks for MOS2, and 9.2 ks for PN. Because of the accelerated build-up of contamination on the 
optical blocking filter of ACIS/{\em Chandra}, the {\em XMM} data are much more sensitive than the {\em Chandra} data below 1.5 keV.
Following Sun et al. (2010) and Ge et al. (2018), the X-rays data have been reduced using the standard procedures 
with CIAO 4.10 and CALDB 4.7.9 ({\em Chandra}), and SAS 17.0.0 ({\em XMM}).

\subsection{MUSE spectroscopy}

The MUSE data have been taken as part of the 097.D-0408(A) and 098.D-0115(A) programs of the AMUSING supernova survey (Galbany et al. 2018).
Observations have been carried out using the Wide Field Mode in the spectral range 4800-9300 \AA~ with a spectral 
resolution of 2.55 \AA, corresponding to $\sim$ 50 km s$^{-1}$ at H$\alpha$ as described in Galbany et al. (2016) and Kruhler et al. (2017). 
The data have been gathered during two different runs, one in May 2016
with four dithered exposures 625 sec long, and centered at $\sim$ 18 arcsec on the east of the nucleus of the 
galaxy, while the second one in January 2017, with again four dithered exposures of 701 sec, centered at $\sim$ 75 
arcsec on the east  of the nucleus. 
To secure a correct sampling of the sky on this extended galaxy, two 
exposures have been taken with the telescope pointed on an empty region close to the galaxy. The observations have been 
carried out in mostly clear sky conditions. The typical sensitivity of MUSE to low and extended surface brightness features at H$\alpha$ 
is $\Sigma(H\alpha)$ $\sim$ 4$\times$10$^{-18}$ erg s$^{-1}$ cm$^{-2}$ arcsec$^{-2}$.


The MUSE data have been reduced using a combination of in-house python procedures and the MUSE pipeline (Weilbacher et al. 2014), as first presented in Fumagalli et al. (2014) and
later improved in Fossati et al. (2016) and Consolandi et al. (2017). These procedures perform the cube reconstruction using the daytime calibrations, and later improve the
illumination spatial uniformity. The skylines subtraction is achieved with the Zurich Atmospheric Purge (ZAP, V1.0, Soto et al. 2016). Individual exposures are projected on a 
common output grid and are then combined with mean statistics. Lastly, the cube is smoothed by 10$\times$10 spaxels in the spatial direction to improve the $S/N$ ratio of the spectra 
without compromising the image quality. 

The stellar continuum is modeled and subtracted from the datacube using the GANDALF code (Sarzi et al. 2006), which uses the penalised pixel-fitting code (pPXF, Cappellari \& Emsellem 2004). We only fit spaxels with a S/N ratio per spectral channel greater than 5 and we used the MILES spectral library (Vazdekis et al. 2010), and we record the
stellar kinematics from the best fit spectra. 
We fit the emission lines from the continuum subtracted cube using the {\sc KUBEVIZ} software as detailed in Fossati et al. (2016). After fitting all the spaxels independently 
we flag as good fits those with a S/N ratio greater than 5 for the H$\alpha$ flux. In this work we also run a new feature of the  {\sc KUBEVIZ} code that iteratively attempts 
to improve the unreliable fits using initial guesses obtained from the nearby good spaxels. With this method we obtain good fits for $\sim 2\%$ more spaxels than with the 
original fitting method. 

Lastly, we find no ionised gas emission  in the eastern datacube and we find that the stellar kinematics fits are highly uncertain in this cube due to the very low surface brightness 
of the stellar continuum. For these reasons we only use the central field in the following analysis.

\subsection{Fabry-Perot spectroscopy}

High resolution 3D spectroscopic observations of NGC 4424 have been carried out using two different Fabry-Perot interferometers,
SAM-FP on the 4.2 m SOAR telescope located at Cerro Pach\'on (Chile), and GHASP on the 1.93 m Observatoire de Haute Provence (OHP) telescope (France). 
SAM-FP, a Fabry-Perot mounted inside the SOAR telescope Adaptive-optics Module, is the only imaging Fabry-Perot interferometer 
which uses a laser-assisted ground layer adaptive optic over a 3 $\times$ 3 arcmin$^2$ \FOV (Mendes de Oliveira et al. 2017).  
SAM-FP is equipped with a single e2v CCD with 4096 $\times$ 4112 pixels with a 2 $\times$ 2 binned pixel of size 0.18 $\times$ 0.18 arcsec$^2$. 
The observations have been carried out in April 2017 during the commissioning of the instrument. 
The galaxy has been observed with a 1 h exposure in dark sky under photometric conditions. Thanks to the Ground Layer Adaptive Optics (GLAO) module, 
the mean image quality is 0.55 arcsec.
GHASP is a focal reducer including a Fabry-Perot providing a 5.8 $\times$ 5.8 arcmin$^2$ \FOV equipped with a 512 $\times$ 512 Imaging Photon 
Counting System (IPCS) with a pixel scale of 0.68 $\times$ 0.68 arcsec$^2$ (Garrido et al. 2005). The galaxy has been observed in dark sky time
during photometric conditions, with a total exposure time of 4 h and a typical seeing of 3 arcsec.
The Free Spectral Range of both Fabry-Perots, 492 km s$^{-1}$ and 378 km s$^{-1}$ for SAM-FP and GHASP, were scanned through 36 and 32 channels, respectively. 
This provides a comparable spectral resolution of $R$ $\simeq$ 12000 at H$\alpha$. For both instruments, suitable 15 \AA\ FWHM interference filters 
have been used to select the spectral range of interest. The Ne I 6598.95 \AA\ emission line was used for wavelength calibration.
The line detection limits for the diffuse emission of both instruments after Voronoi binning is $\sim$ 2.5 $\times$ 10$^{-17}$ erg s$^{-1}$ cm$^{-2}$ arcsec$^{-2}$
at OHP and $\sim$ 5 $\times$ 10$^{-17}$ erg s$^{-1}$ cm$^{-2}$ arcsec$^{-2}$ at Cerro Pach\'on.

 
Fabry-Perot data have been reduced following the standard procedures fully described in Daigle et al. (2006) and Epinat et al. (2008). 
Night-sky line subtraction was applied after data cube construction and wavelength calibrations. 
Using adaptive binning techniques based on 2D-Voronoi tessellations applied to the 3D data cubes, we derived Voronoi H$\alpha$ data cubes 
from which, among others, H$\alpha$ maps, radial velocity and velocity dispersion fields are computed.  
With the spatial adaptive binning technique, a bin is accreting new pixels until it reaches an \emph{a priori} criterion, which has been fixed here 
to a $S/N$ = 7 for the emission line within the bin for both sets of observations.   
The main advantage of this technique is that the best spatial resolution is obtained at each point of the galaxy and that 
no spatial contamination between the bins is possible, each bin corresponds to the same $S/N$ and therefore has the same statistical weight.

\section{Multifrequency data}

\subsection{The data}

The galaxy NGC 4424 is located close to the core of Virgo and has been observed during several multifrequency
surveys of the cluster. Deep GALEX ultraviolet images in the $FUV$ ($\lambda_c$ = 1539 \AA; integration time = 2008 sec) and 
$NUV$ ($\lambda_c$ = 2316 \AA; integration time = 2009 sec)
bands have been gathered during the GUViCS survey of the Virgo cluster (Boselli et al. 2011). The galaxy has been also observed in the $ugiz$
broad-bands during the NGVS survey (Ferrarese et al. 2012), in the near-infrared $H$-band by Boselli et al. (2000), in the mid-infrared
by WISE (Wright et al. 2010) and by \textit{Spitzer} in the four IRAC bands (Ciesla et al. 2014), 
and in the far-infrared still by \textit{Spitzer} with MIPS at 24, 70, and 160 $\mu$m (Bendo et al. 2012) and by \textit{Herschel} 
with PACS at 100 and 160 $\mu$m (Cortese et al. 2014) and SPIRE at 250, 350, and 500 $\mu$m (Ciesla et al. 2012) as part of the \textit{Herschel}
Reference Survey (HRS; Boselli et al. 2010) and the \textit{Herschel} Virgo Cluster Survey (HeViCS; Davies et al. 2010). 
The images of the galaxy in different photometric bands, including the continuum-subtracted NB H$\alpha$, the $FUV$, $NUV$, $u$,
IRAC1 3.6 $\mu$m, and the MIPS 24 $\mu$m bands, are shown in Fig. \ref{multifrequency}. High quality images of the
galaxy in the visible have been also obtained with the WFC3 on the HST by Silverman et al. 2012 (Fig. \ref{HST}). 
HI VLA, KAT-7, and WSRT data of NGC 4424 are available from Chung et al. (2009) and Sorgho et al. (2017). 
VLA radio continuum data at 6 cm and 20 cm, including polarisation,
are also available (Vollmer et al. 2013).

   \begin{figure*}
   \centering
   \includegraphics[width=1\textwidth]{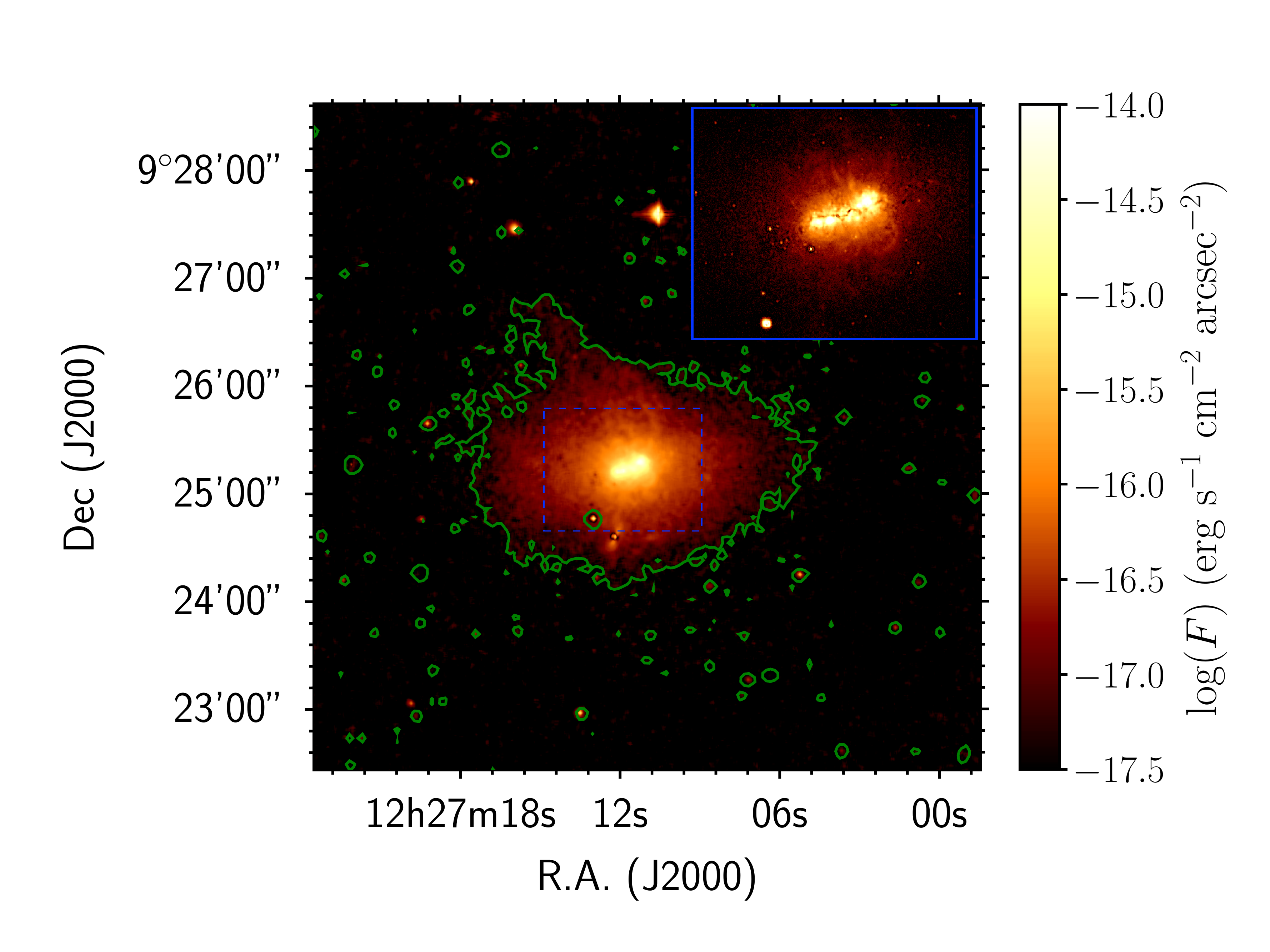}
   \caption{Continuum-subtracted H$\alpha$ image of NGC 4424
   with contours at $\Sigma(H\alpha)$ = 4$\times$10$^{-18}$ erg s$^{-1}$ cm$^{-2}$ arcsec$^{-2}$. A prominent tail is 
   visible in the northern-east direction extending from the ionised gas halo. The inset shows a contrasted image of the central regions (here on an arbitrary surface brightness scale), 
   with prominent filaments of ionised gas.
 }
   \label{Ha}%
   \end{figure*}


  \begin{figure*}
   \centering
   \includegraphics[width=1\textwidth]{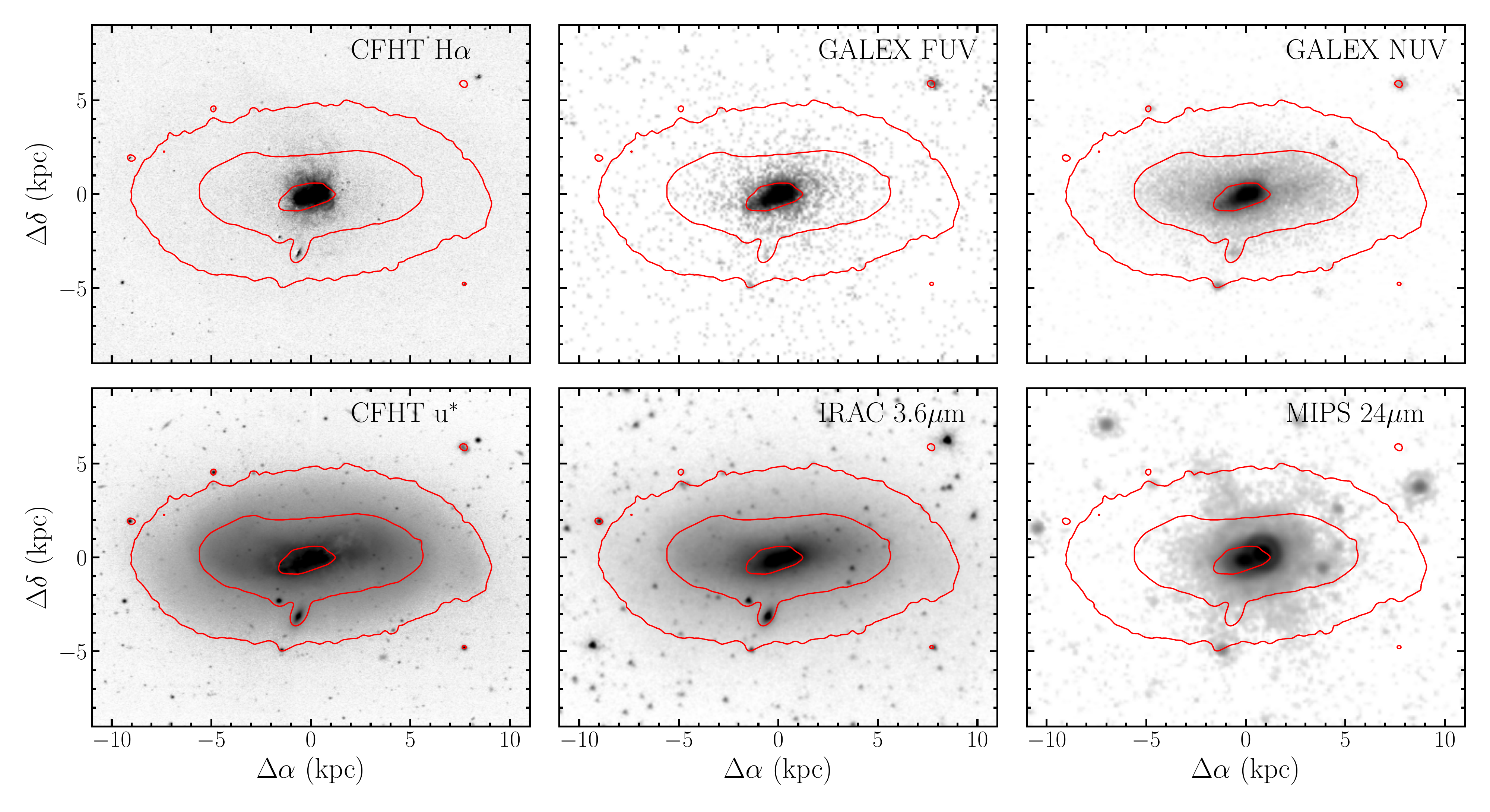} 
   \caption{Multifrequency images on the same scale of the galaxy NGC 4424. Upper panels, from left to right: continuum-subtracted H$\alpha$ (VESTIGE), $FUV$ (GALEX), $NUV$ (GALEX); lower
    panels: $u$ (NGVS), IRAC 3.6 $\mu$m ($\textit{Spitzer}$), MIPS 24 $\mu$m ($\textit{Spitzer}$). Red contours are drawn from the $u$-band image at 22nd, 24th, and 26th mag arcsec$^{-2}$
    and highlight the truncation of the disc at different wavelengths.
 }
   \label{multifrequency}%
   \end{figure*}

   \begin{figure}
   \centering
   \includegraphics[width=9cm]{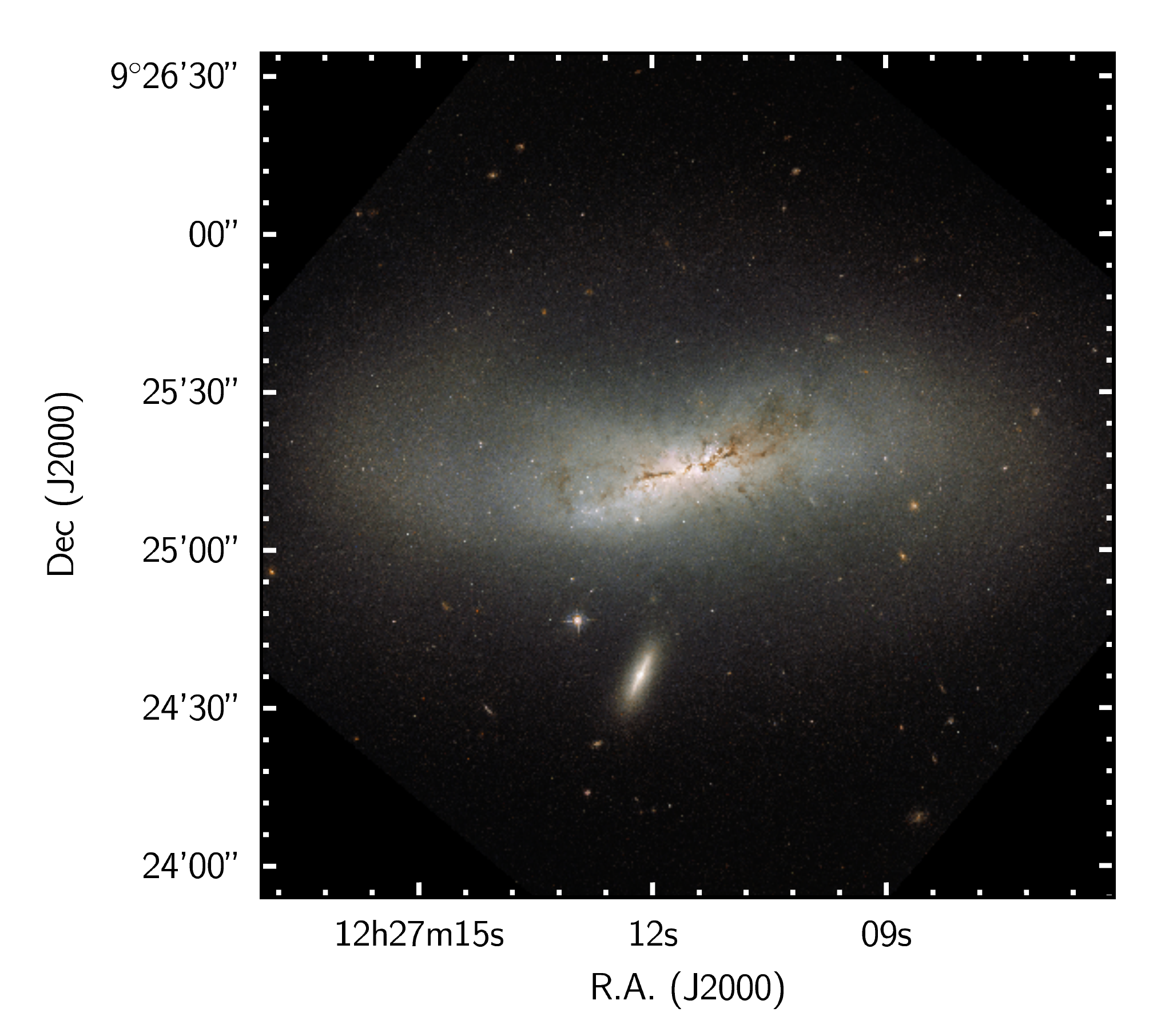}\\
   \caption{HST colour image of the galaxy NGC 4424, showing at an unequaled sensitivity and angular resolution a perturbed morphology, with plumes and filaments of dust in absorption
   located along the major axis of the galaxy or perpendicular to it. 
 }
   \label{HST}%
   \end{figure}



\subsection{Image analysis}

A first glance to the new images presented in this paper clearly underlines the peculiar morphology of the galaxy at all wavelengths 
although with some remarkable differences with respect to what found in previous
works. First of all, the very deep NGVS images (Figs. \ref{NGVS} and \ref{multifrequency}) do not show the presence of any 
fine structure in the stellar component (shells, rings, plumes, tails) in the outer regions
generally interpreted as a proof of a recent merging event (e.g. Duc et al. 2011). We recall that these NGVS images have been gathered and reduced using Elixir-LSB,
a pipeline expecially tailored to detect extended low surface brightness features (Ferrarese et al. 2012, Paudel et al. 2017). 
Fine structures in the stellar component are not detected down to a surface brightness limit of $\mu_g$ $\simeq$ 29 mag arcsec$^{-2}$. 
The excellent HST image of the inner region 
not only confirms a perturbed morphology, but also shows plumes and filaments of dust in absorption located along the major axis of the galaxy
or perpendicular to it (Fig. \ref{HST}). A peculiar distribution of the dust emission is also revealed for the first time by the 
24 $\mu$m MIPS/\textit{Spitzer} image (Fig. \ref{multifrequency}). This figure shows a $\simeq$ 5 kpc  
filament (projected length) extending from the nucleus to the north north-east. 
The 3.6 $\mu$m IRAC/\textit{Spitzer} image (Fig. \ref{multifrequency}), sensitive to the distribution of the old stellar population, 
shows an elongated feature slightly misaligned with respect to the major axis, with a prominent compact structure at the western edge and another one in the centre.


Thanks to their excellent quality in terms of sensitivity and angular resolution 
the very deep VESTIGE 
continuum-subtracted NB H$\alpha$ image shows several features unobserved so far, namely:\\
1) a spectacular tail of ionised gas in the north-east $\simeq$ 10 kpc long (projected length) extending from the central star forming region (Figs. \ref{NGVS} and \ref{Ha}). 
This feature, which corresponds to the one detected at 24 $\mu$m (Fig. \ref{multifrequency}) and 
has a typical surface brightness of $\Sigma (H\alpha)$ $\simeq$ 4 $\times$ 10$^{-18}$ erg s$^{-1}$ cm$^{-2}$ arcsec$^{-2}$, is diffuse and does not contain 
clumpy spots indicating the presence of HII regions. 
The star formation activity is present only in the central $\simeq$ 3 kpc$^2$
region, and is mainly located along the thin elongated structure visible in the near-infrared, and in particular at its edges. There are, however, a few blobs 
of star formation south to the nucleus outside the stellar disc. \\
2) filaments of ionised gas perpendicular to the star forming bar extending up to $\simeq$ 1.5 kpc. \\
3) a low-surface brightness ($\Sigma (H\alpha)$ $\simeq$ 4 $\times$ 10$^{-18}$ erg s$^{-1}$ cm$^{-2}$ arcsec$^{-2}$) diffuse halo extending along the major axis of the galaxy
up to $\simeq$ 6-7 kpc from the nucleus. The lack of any 
structured feature suggests that this emission is due to the diffuse gas ionised by the radiation escaping from the nuclear starburst or that it is gas shock-ionised by the ram pressure 
stripping event (see Sect. 4.1.2). The activity of star formation in the outer disc is thus completely quenched.

\section{Physical parameters}

\subsection{Narrow-band imaging}

\subsubsection{The galaxy}

The VESTIGE data can be used to estimate the total H$\alpha$ luminosity of the galaxy and of the tail of ionised gas detected in the north-eastern direction.
The total H$\alpha$ flux of the galaxy including its extended tail is log $f(H\alpha+[NII])$ = -12.12 $\pm$ 0.02 erg s$^{-1}$ cm$^{-2}$, where the uncertainty on the flux has been
measured after deriving the uncertainty from the sky background on an aperture similar to the size of the galaxy in 100 regions randomly located
on the frame around the target. Assuming a typical [NII]$\lambda$6583\AA/H$\alpha$ $=$ 0.36 and a Balmer decrement H$\alpha$/H$\beta$ $=$ 4.45
as measured from the MUSE data in the central field (see Sect. 4.3), the total extinction-corrected H$\alpha$ luminosity of the galaxy is 
$L(H\alpha)$ = 4.50 ($\pm$ 0.21) $\times$ 10$^{40}$ erg s$^{-1}$, which corresponds to a $SFR$ = 0.25 $\pm$ 0.04 M$_{\odot}$yr$^{-1}$ assuming the escape fraction and the fraction of ionising
photons absorbed by dust before ionising the gas to be zero (Boselli et al. 2009), the Kennicutt (1998) $SFR$-$L(H\alpha)$ calibration, and a Chabrier (2003) IMF.

\subsubsection{The diffuse tail}

We also measured the total H$\alpha$ emission in the tail of ionised gas within a box of size 77 $\times$ 32 arcsec$^2$, P.A.=-30 deg., centered at R.A. = 12:27:13.56;
Dec. = +9:26:17.1 obtaining log $f(H\alpha+[NII])$ = -13.85 $\pm$ 0.19 erg s$^{-1}$ cm$^{-2}$. The lack of IFU data in this low surface brightness region prevents us from
having an accurate estimate of the [NII] contribution and of the Balmer decrement. The BPT diagram analysed in Sect. 4.3 indicates that [NII]/H$\alpha$ ratio increases 
in the outer regions of the galaxy where it reaches [NII]$\lambda$6583\AA/H$\alpha$ $\simeq$ 1.0 probably because of the presence of shock-ionised gas. 
The MUSE map of the Balmer decrement indicates that in these outer regions H$\alpha$/H$\beta$ $\simeq$ 4.5 ($A(H\alpha)$ $\simeq$ 1.2 mag). Assuming that these ratios are also valid 
in the diffuse extraplanar tail, where the gas is supposedly shock-ionised and dust might be present as indicated by the extended 24 $\mu$m emission, 
we estimate that the total H$\alpha$ luminosity of the northern diffuse component is $L(H\alpha)$ $\simeq$ 5.34 ($\pm$ 2.34) $\times$ 10$^{38}$ erg s$^{-1}$. 
Using relation (2) in Boselli et al. (2016a), and assuming a filling factor $f$ = 0.1 and a gas distributed within a cylinder of diameter ($\simeq$ 2.6 kpc) and height ($\simeq$ 7 kpc)
similar to the deprojected rectangular box used for the flux extraction in the tail, we get a rough estimate of the typical gas density in the tail, $n_e$ $\simeq$ 7 $\times$ 10$^{-2}$ cm$^{-3}$,
with a corresponding total ionised gas mass $M_{tail}(H\alpha)$ $\simeq$ 6.0 $\times$ 10$^6$ M$_{\odot}$, to be compared to the HI mass
in the tail, $M_{tail}(HI)$ $\simeq$  5.1 $\times$ 10$^7$ M$_{\odot}$ (Sorgho et al. 2017). We can also estimate the typical recombination time scale for the diffuse gas,
using eq. (5) in Boselli et al. (2016a), $t_{rec}$ $\simeq$ 1.4 Myrs.


\subsection{X-rays imaging}

\begin{figure}
   \centering
   \includegraphics[width=9cm]{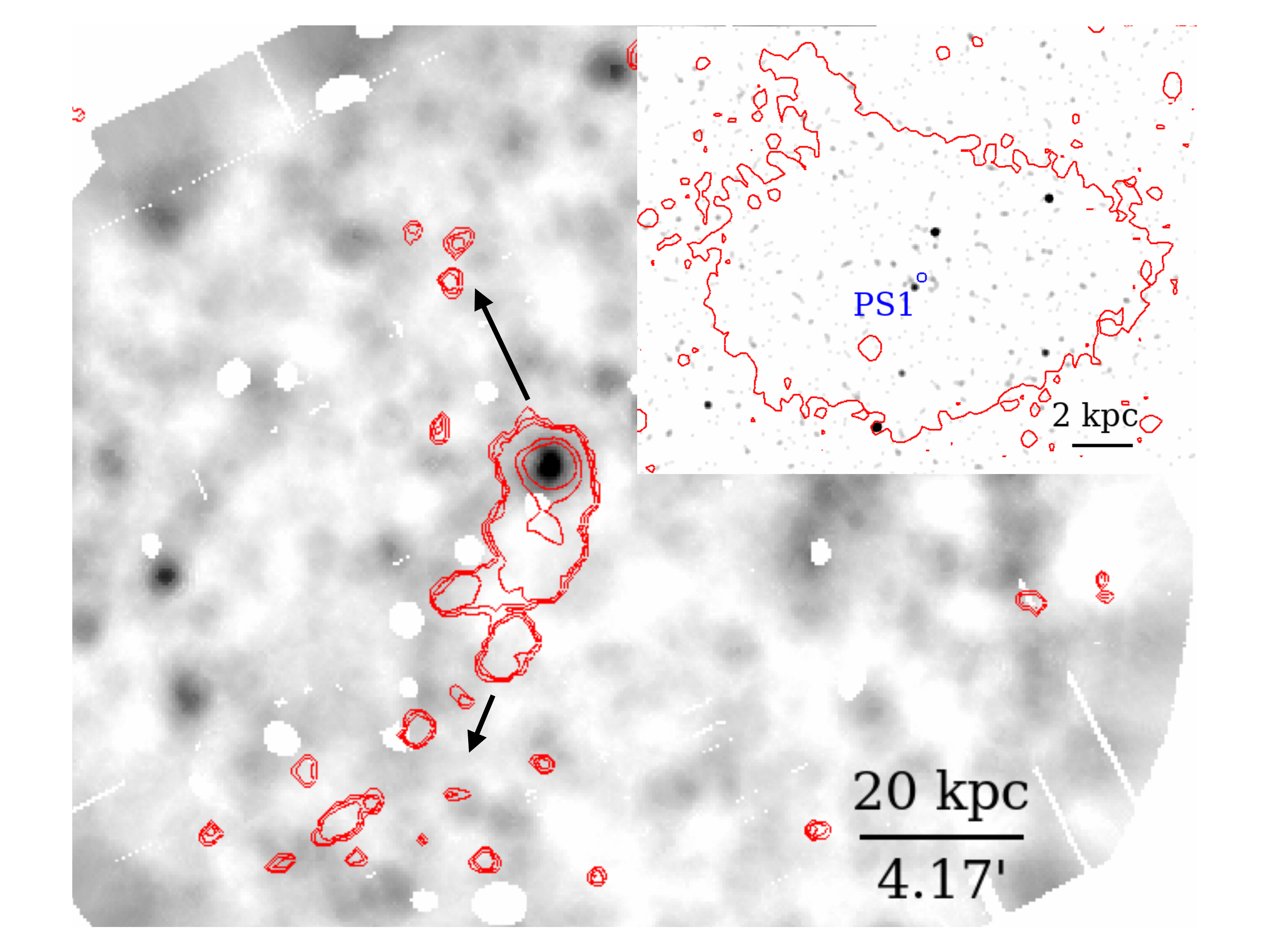}
   \caption{The adaptively smoothed 0.4 - 1.3 keV {\em XMM} image of the NGC 4424 with the HI contours (from Chung et al. 2009) superposed. 
   Bright X-ray point sources outside of the galaxy are masked. No soft X-ray enhancement is observed in the position of the HI tail. 
   The two arrows at the north and at the south of the galaxy point towords the centre of Virgo cluster A and B, respectively. Their length
   is proportional to the distance to M87 and M49.
   The zoom-in region on the upper right shows the {\em Chandra} 0.5 - 7 keV image of the galaxy, with the H$\alpha$ contours (as in Fig. \ref{Ha}) superposed. 
   The small blue circle with a radius of 2$''$ shows the putative nuclear position, without a corresponding X-ray source. 
   An obscured X-ray source 4.9$''$ to the southeast of the nucleus (PS1) is detected by {\em Chandra}.
 }
   \label{XMM}%
   \end{figure}

Fig. \ref{XMM} shows the {\em XMM} and {\em Chandra} images of NGC 4424.
The putative nucleus is undetected by {\em Chandra}. Assuming a power law
model with a photon index of 1.7, the 3$\sigma$ upper limit of the
2 - 10 keV luminosity is 5.6 $\times$ 10$^{37}$ erg s$^{-1}$ for an absorption
column density of 1.7 $\times$ 10$^{20}$ cm$^{-2}$ (Galactic absorption).
As a significant amount of intrinsic absorption is expected in NGC 4424,
the 3$\sigma$ upper limit becomes 6.6 $\times$ 10$^{37}$ erg s$^{-1}$
and 1.5 $\times$ 10$^{38}$ erg s$^{-1}$ for an absorption column density of 
10 times and 100 times greater than the Galactic value, respectively.
The expected super massive black hole (SMBH) mass of NGC 4424 derived using the Marconi \& Hunt (2003) relation is $\sim$ 2.4 $\times$ 10$^{7}$ M$_{\odot}$.
Thus, in NGC 4424 the SMBH is quiescent and has an Eddington ratio of $< 10^{-6}$.
There is indeed an X-ray source 4.9$''$ to the southeast of the nucleus (PS1 in Fig. \ref{XMM}).
It is only detected in 2-7 keV, suggesting it is obscured. Its X-ray hardness ratio
is consistent with an absorption column density greater than 2.5 $\times$ 10$^{22}$ cm$^{-2}$.
The 2 - 10 keV luminosity is 4.3 $\times$ 10$^{38}$ erg s$^{-1}$ for an absorption column density
of 2.5 $\times$ 10$^{22}$ cm$^{-2}$ and 1.1 $\times$ 10$^{39}$ erg s$^{-1}$ for an
absorption column density of 10$^{23}$ cm$^{-2}$. If it is associated with NGC 4424,
this point source can be an X-ray binary (almost an ultraluminous X-ray source in that case)
or the nucleus of a merged galaxy.

As shown in Fig. \ref{XMM}, the HI tail is undetected in X-rays.
The 3$\sigma$ upper limit of the 0.4 - 1.3 keV emission in the HI tail can be estimated from
the {\em XMM} data, and is 4.1$\times10^{-16}$ erg cm$^{-2}$ s$^{-1}$ arcmin$^{-2}$ and
4.8$\times10^{-16}$ erg cm$^{-2}$ s$^{-1}$ arcmin$^{-2}$ for an assumed gas temperature of 0.6 keV and 0.3 keV, respectively.
The resulting X-ray bolometric luminosity of the tail is thus at least 370 - 700 times smaller than
that of ESO 137-001 (Sun et al. 2010). Although the mixing of the cold gas stripped
from the galaxy with the hot ICM enhances the soft X-rays emission within the tail, 
the ambient pressure, which in the position of NGC 4424 is $\sim$ 100 smaller than around ESO 137-001
(see Sect. 5.4), is not sufficient to make the tail detectable with the existing X-rays observations (Tonnesen et al. 2011).

\subsection{Spectroscopy}

\subsubsection{Line ratios}

\begin{figure}
   \centering
   \includegraphics[width=9cm]{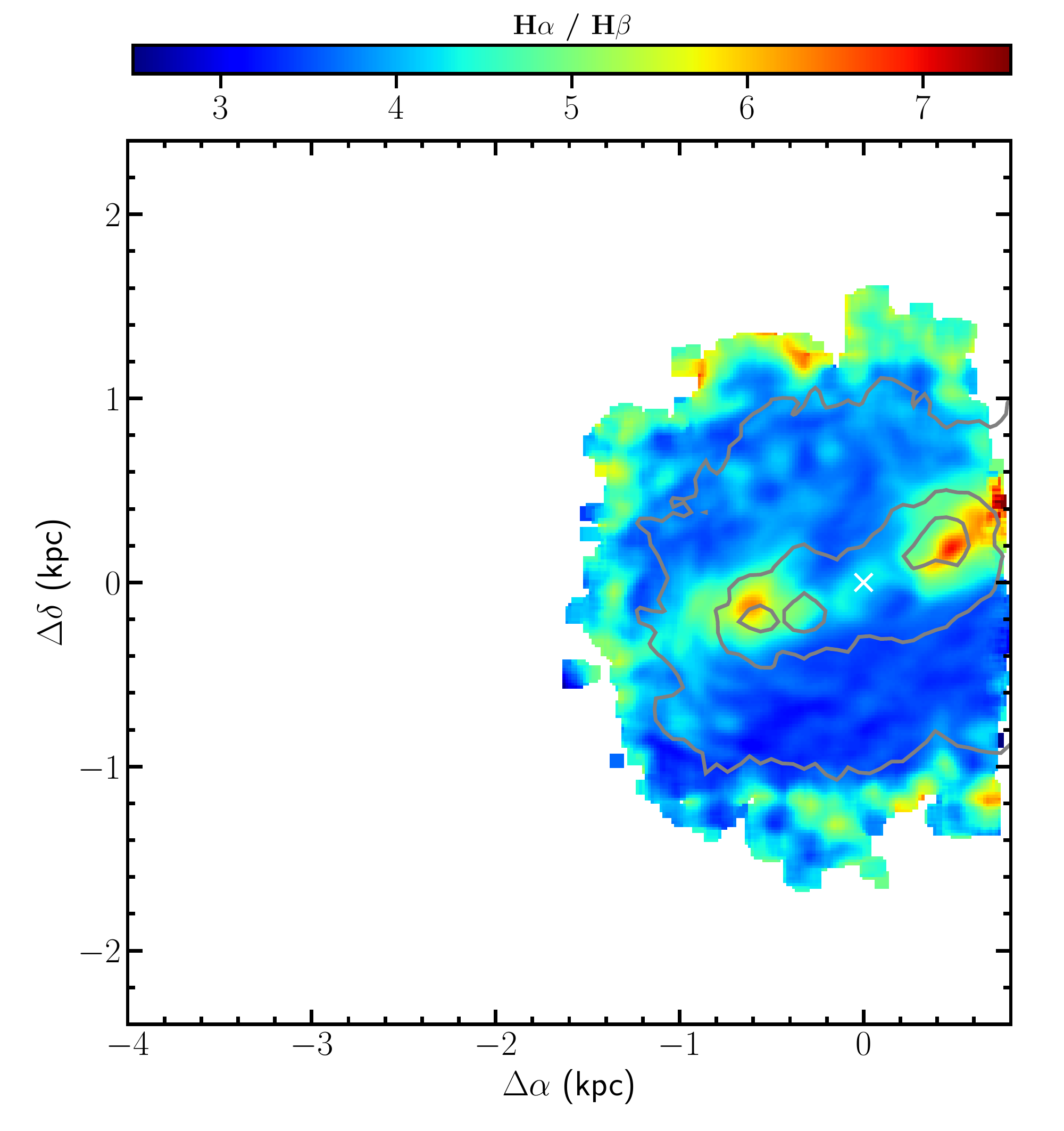}
   \caption{The spaxel distribution of the Balmer decrement H$\alpha$/H$\beta$ derived from the MUSE pixels with a signal-to-noise ratio $S/N$ $>$ 5.
   The cross indicates the position of the nucleus, the grey contours the distribution of the H$\alpha$ emitting gas (contour levels are $\Sigma(H\alpha)$ = 5 $\times$ 10$^{-17}$ - 10$^{-16}$
   erg s$^{-1}$ cm$^{-2}$ arcsec$^{-2}$).
 }
   \label{BDM}%
   \end{figure}

The MUSE data can be used to derive the typical properties of the ISM. The Balmer decrement shown in Fig. \ref{BDM} traces 
the dust attenuation in the Balmer lines. The H$\alpha$/H$\beta$ ratio ranges from $\simeq$ 7 in the western peak of H$\alpha$ emission ($A(H\alpha)$ $\simeq$ 2.3 mag) to $\simeq$ 3
in the diffuse gas ($A(H\alpha)$ $\simeq$ 0.1 mag). It is worth noticing, the eastern peak of H$\alpha$ emission does not match with the peak of dust attenuation which is 
located further out ($\simeq$ 0.3 kpc) along the elongated structure dominating the ionised gas emission.

The [SII]$\lambda$6716/6731 line ratio, a direct tracer of the gas density, is always fairly high, 
with the lower values found on the two bright H$\alpha$ nuclei, where [SII]$\lambda$6716/6731 $\simeq$ 1.2-1.3, corresponding to gas densities
$n_e$ $\simeq$ 150 cm$^{-3}$, while [SII]$\lambda$6716/6731 $\simeq$ 1.4-1.5 in the diffuse gas ($n_e$ $\lesssim$ 30 cm$^{-3}$) (Osterbrock \& Ferland 2006; Proxauf et al. 2014).  
Line measurements can also be used to derive the metallicity of the gas. Using the calibration of Curti et al. (2017) based on the [OIII], H$\beta$, [NII] and H$\alpha$ lines, 
we obtain typical metallicities
ranging from 12 $+$ log O/H $\simeq$ 8.65 in the outer gas to 12 $+$ log O/H $\simeq$ 8.8 in the inner regions. 
These values are comparable to those derived for galaxies of similar stellar mass (e.g. Hughes \& Cortese 2013).



\begin{figure*}
   \centering
   \includegraphics[width=18cm]{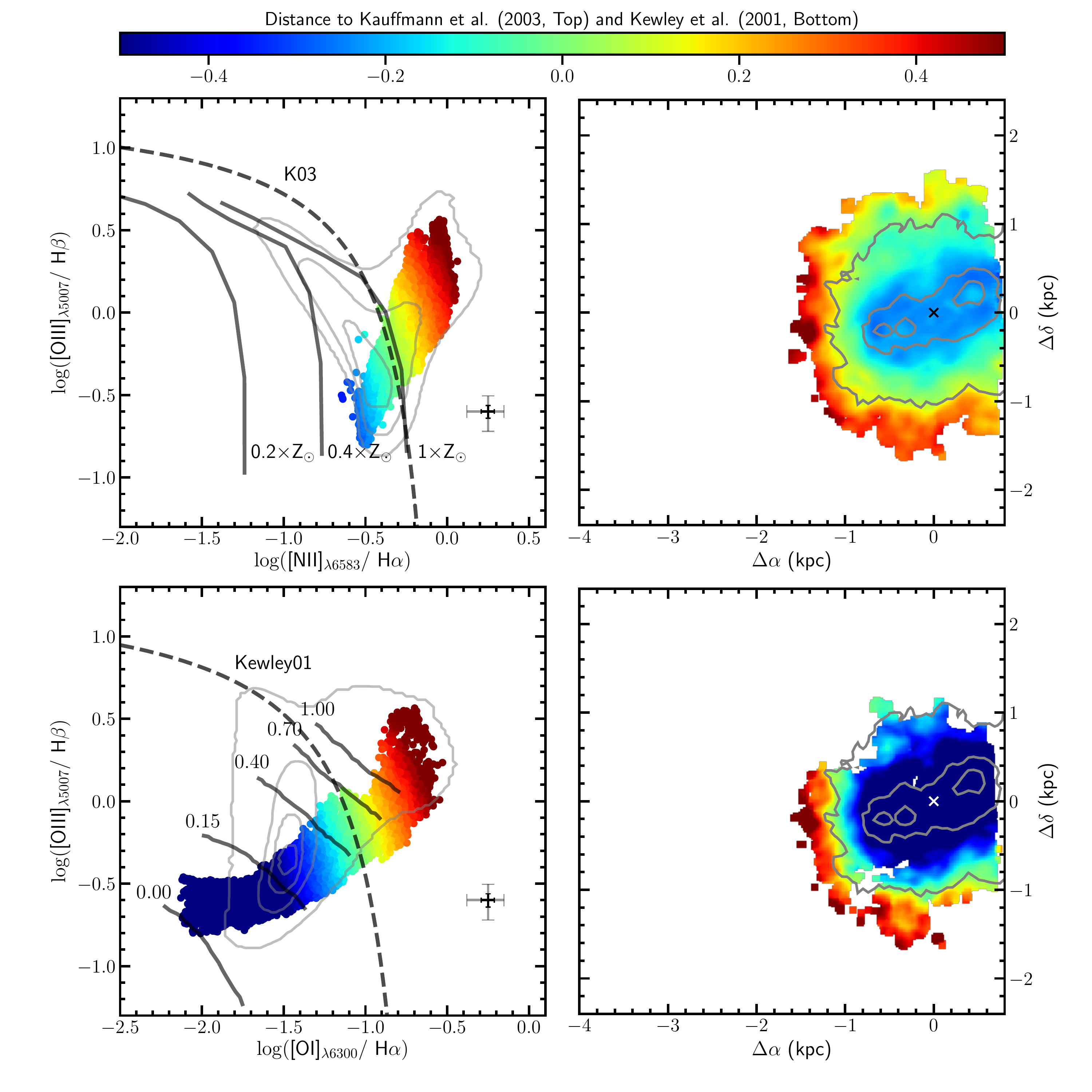}
   \caption{The line diagnostic BPT diagrams log([OIII]$_{\lambda 5007}$/H$\beta$) vs. log ([NII]$_{\lambda  6583}$/H$\alpha$) (upper left) 
   and log([OIII]$_{\lambda 5007}$/H$\beta$) vs. log ([OI]$_{\lambda  6300}$/H$\alpha$) (lower left) derived from those MUSE pixels with a signal-to-noise ratio $S/N$ $>$ 5. 
   The different points are colour-coded according to their distance
   from the dashed lines defined by Kauffman et al. (2003) (upper panel) and Kewley et al. (2001) (lower panel) to separate AGN from star forming regions. The grey
   contours shows the distribution of a random sample of nuclear spectra of SDSS galaxies in the redshift range 0.01 - 1 and stellar mass 10$^9$ - 10$^{11}$. The 
   solid thick lines in the upper left panel show the expected behaviour of star forming regions as derived from the photo-ionisation models of Kewley et al. (2001)
   for three different metallicities (0.2, 0.4, 1 $Z_{\odot}$). Those in the lower left panel show the shock models of Rich et al. (2011) for increasing shock fraction
   (from left to right) in a twice solar gas.
   The right panels show the distribution of the same points colour-coded as in the left panels 
   over the disc of the galaxy. The cross shows the centre of the galaxy, the grey contours the distribution of the H$\alpha$ emitting gas.
 }
   \label{BPT}%
   \end{figure*}

The spectroscopic data can also be used to derive several line diagnostic diagrams such as those proposed by Baldwin et al. (1981; BPT diagrams). These diagrams, shown in 
Fig. \ref{BPT}, can be used to identify the dominant ionising source of the gas. 
Figure \ref{BPT} shows that the gas is photo-ionised by the young stars in the inner regions, while 
it is shock-ionised in the outer ones. The high cut in the signal-to-noise in all the emission lines used in the two diagrams ($S/N$ $>$ 5)
and the agreement in the [OIII]/H$\beta$ vs. [NII]/H$\alpha$, [OIII]/H$\beta$ vs. [OI]/H$\alpha$, and [OIII]/H$\beta$ vs. [SII]/H$\alpha$
(not shown here) BPT diagrams secure this result against systematic effects in the data. Figures \ref{BPT} and \ref{XMM} do not show the presence of any hard ionising source near the nucleus,
suggesting a very low accretion rate despite the abundant cold material in the nuclear region (Cortes et al. 2006).

\subsubsection{Stellar Kinematics}

Figure \ref{velMUSE} shows the stellar kinematics derived using the MUSE data (lower left panel). The stellar velocity field of NGC 4424, although showing a gradient from the nucleus to the 
eastern regions\footnote{The velocity does not seem to increase in the eastern disc mapped by the second field of MUSE. The low $S/N$ level 
in this outer region prevents us from deriving an accurate stellar velocity field.} of $\simeq$ 40 km s$^{-1}$, does not have the typical shape of a rotating disc for a spiral galaxy inclined $\sim$ 60 deg on the 
plane of the sky, with a steep gradient in the inner regions. Furthermore, this rotation ($vel$ $\simeq$ 85 km s$^{-1}$ when corrected for inclination) 
is significantly smaller than the one expected for a rotating system of similar stellar mass ($M_{star}$ $\simeq$ 10$^{10.17}$ $M_{\odot}$;
$vel$ $\simeq$ 200 km s$^{-1}$)
as indeed noticed by Kenney et al. (1996), Rubin et al. (1999), Coccato et al. (2005), and Cortes et al. (2015).
The superior quality of the MUSE data allows us to reconstruct the stellar velocity field down 
to one arcsec angular resolution ($\simeq$ 80 pc). This stellar velocity field shows a kinematically decoupled structure on the elongated feature emitting in H$\alpha$ and in 
the nucleus of the galaxy never seen in previous data. This feature suggests that 
the stellar component associated to the nuclear starburst has a different origin than the outer disc.    

\subsubsection{Gas Kinematics}

\begin{figure*}
   \centering
   \includegraphics[width=0.9\textwidth]{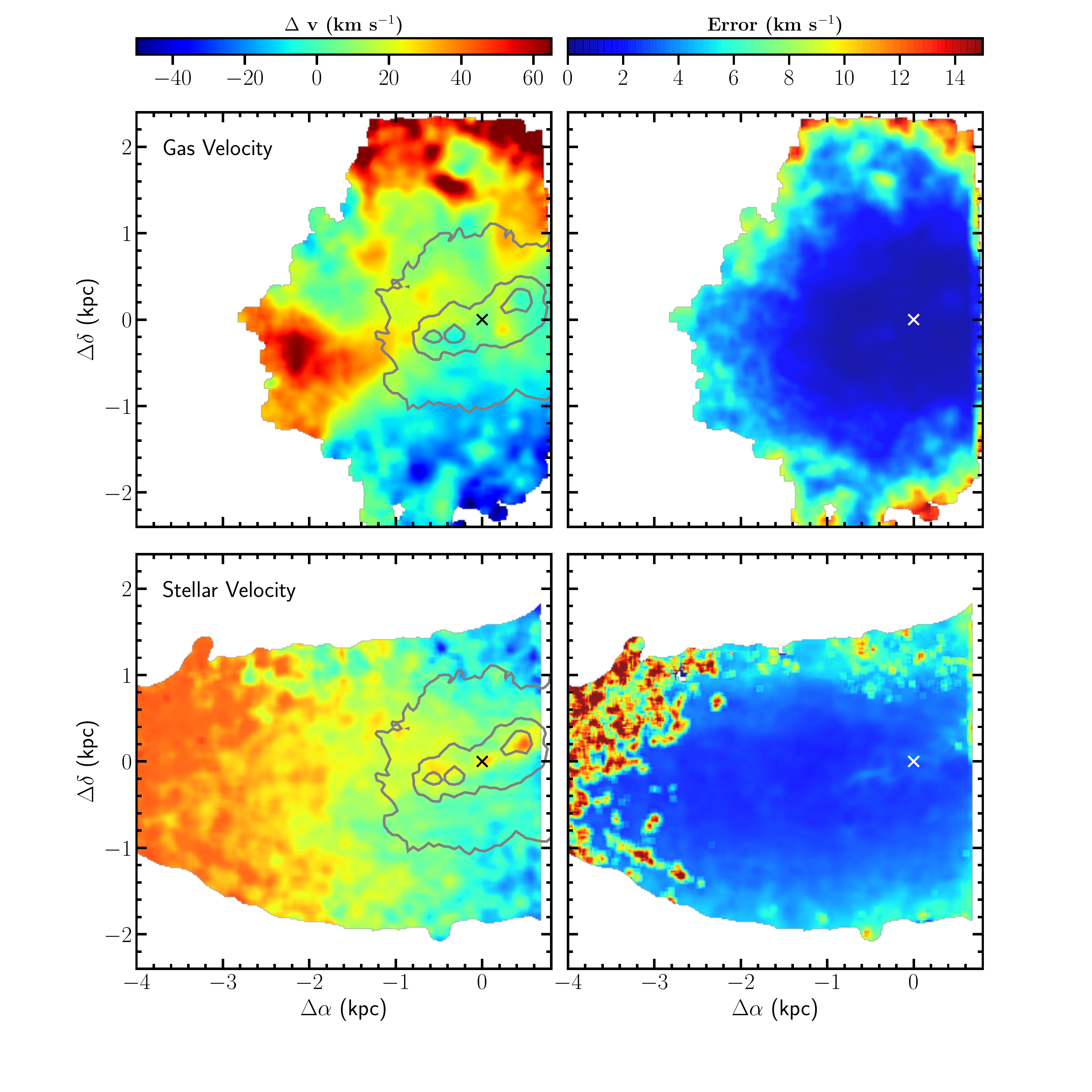}
   \caption{The velocity field (left panels) and the error maps (right panels) of the gaseous (upper panels) and of the stellar (lower panels) components
   derived from the MUSE data.
   The black and white crosses indicate the position of the nucleus, the grey contours the distribution of the H$\alpha$ emitting gas. 
 }
   \label{velMUSE}%
   \end{figure*}

\begin{figure*}
   \centering
   \includegraphics[width=0.9\textwidth]{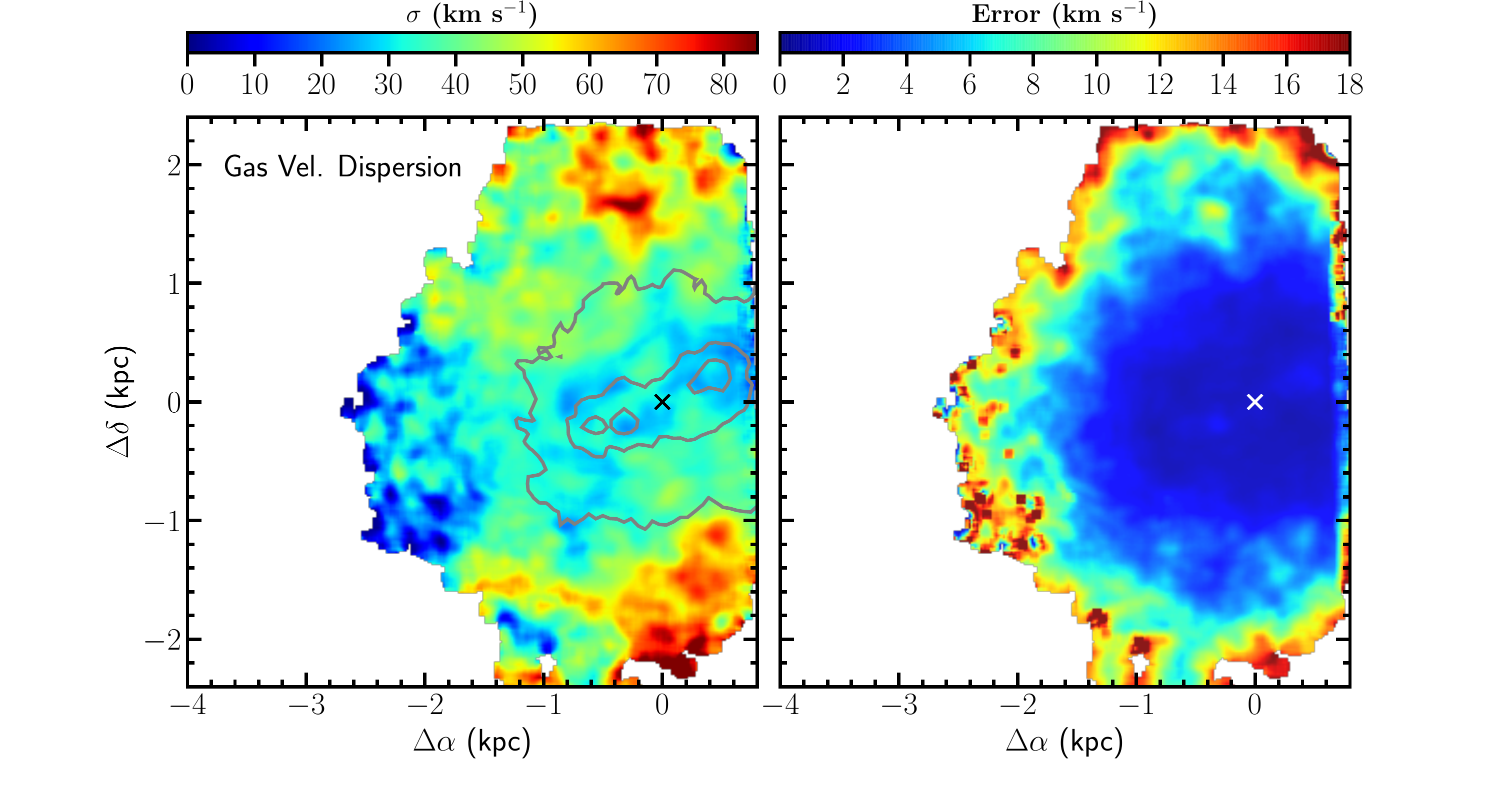}
   \caption{The velocity dispersion (left panels) and the error maps (right panels) of the gaseous component derived from the MUSE data.
   The cross indicates the position of the nucleus, the grey contours the distribution of the H$\alpha$ emitting gas. 
 }
   \label{sigMUSE}%
   \end{figure*}

\begin{figure*}
   \centering
   \includegraphics[width=0.9\textwidth]{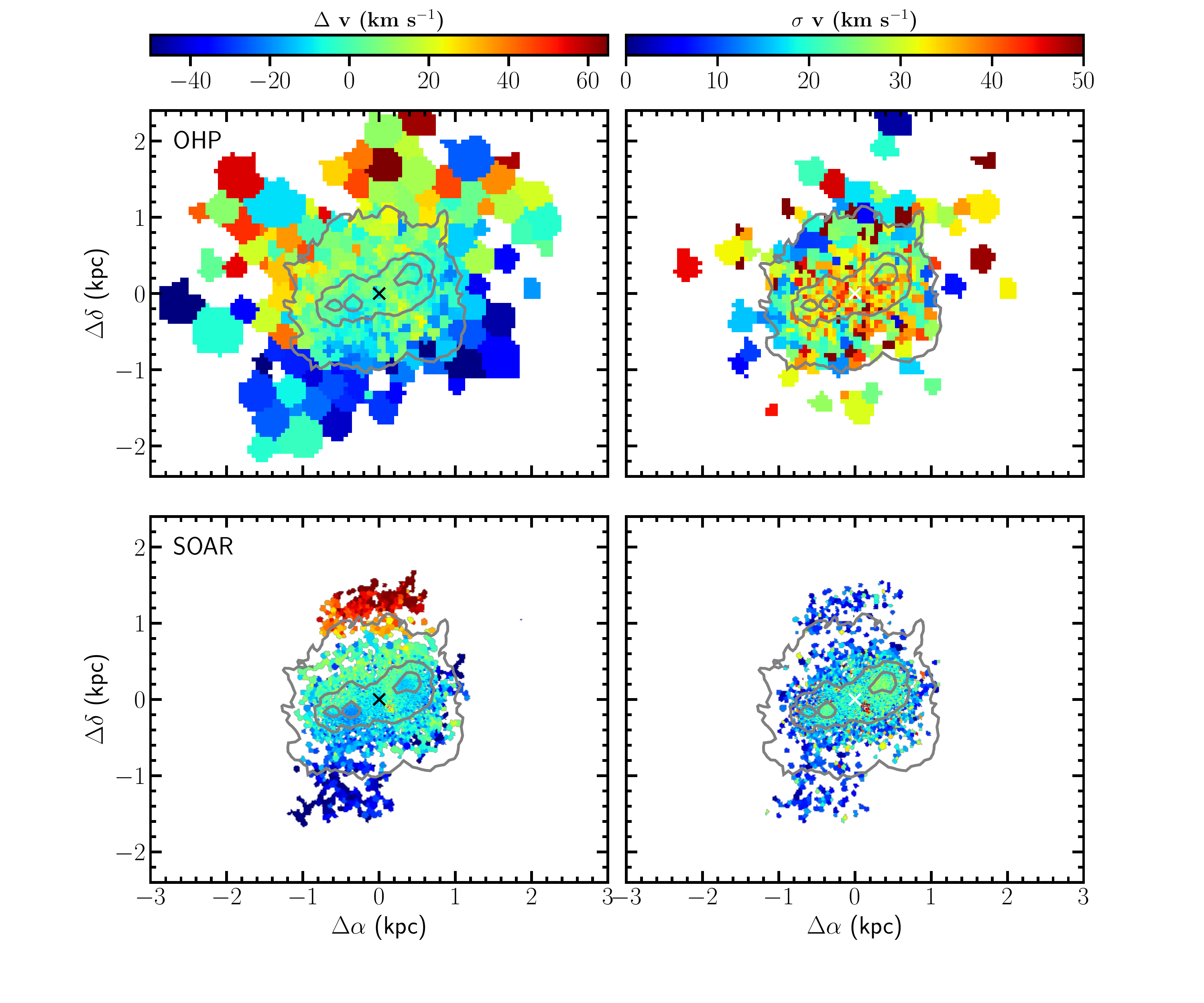}
   \caption{The velocity field (left panels) and the velocity dispersion (right panels) of the gaseous component
   as derived from the OHP (upper panels) and SOAR (lower panels) Fabry-Perot observations.
   The black and white crosses indicate the position of the nucleus, the grey contours the distribution of the H$\alpha$ emitting gas. 
 }
   \label{FP}%
   \end{figure*}

Figure \ref{velMUSE} also shows the velocity of the gas derived using the MUSE data (upper panel), while its velocity dispersion is shown in Fig. \ref{sigMUSE}.
The MUSE data suggest that the gas is rotating at large scales along the major axis as the stellar component, albeit with a slightly steeper gradient. We expect that the same rotation of both components is 
present on the western side where MUSE data are unfortunately unavailable, as indeed suggested by the CO velocity field of Cortes et al. (2006). As for the stellar component, in the inner region we do not see any steep gradient 
in the gas component along the major axis, as expected in a massive unperturbed system.
The two dominant H$\alpha$ emitting regions located at the east and west of the stellar nucleus are at approximately the same line-of-sight velocity ($\simeq$ -10 km s$^{-1}$, see Fig. \ref{FP}),
suggesting that the elongated structure to which they belong is not a rotating bar. Between these two H$\alpha$ emitting regions there is a rather 
narrow (50 pc) north/south filament in the velocity field displaying a higher mean \los velocity ($\simeq$ 30$\pm$10 km s$^{-1}$) surrounded by velocities 
closer to those of the dominant H$\alpha$ blobs than to that of the filament. This filament is also present in the MUSE data. 
The kinematical data suggest that this high velocity filament might be in a different plane than the low velocity component embedding the two dominant H$\alpha$ emitting regions.

Figures \ref{velMUSE} and \ref{FP} consistently indicate that there is a systematic difference in velocity between the northern ($\simeq$ + 50 km s$^{-1}$)
and the southern regions ($\simeq$ - 30 km s$^{-1}$), suggesting the presence of a second gas component.
Figure \ref{sigMUSE} also indicates that in these outer regions the velocity dispersion of the gas is fairly high ($\simeq$ 80 km s$^{-1}$).
This second component might be gas rotating along the minor axis (Coccato et al. 2005), as well as gas infalling 
or outflowing from the nucleus, depending on the orientation of the galaxy disc on the plane of the sky. The line-of-sight overlapping of the minor axis component on the major axis gas could explain why in the inner regions
we do not observe any velocity gradient along the major axis. The high spectral resolution Fabry-Perot data ($\simeq$ 5 km s$^{-1}$) do not show multiple profiles, suggesting
that the two gaseous components are well mixed within this inner region.

\section{Discussion}

\subsection{SED fitting}

\begin{figure*}
   \centering
   \includegraphics[width=0.9\textwidth]{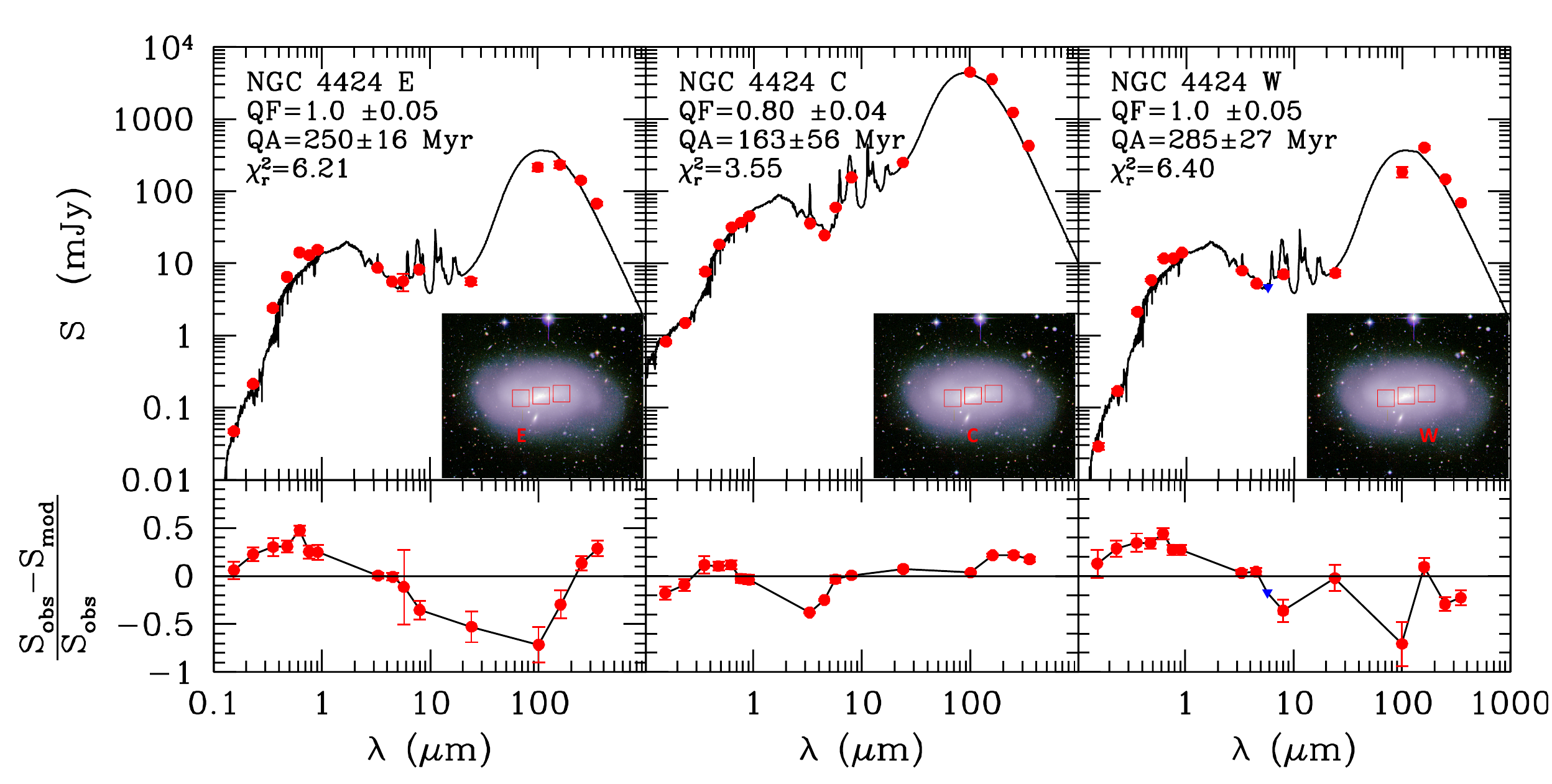}\\
    \caption{The far UV-to-farIR SED of the three selected regions along the disc of NGC 4424 (from east to west).
    The observational data and their error bars are marked with a red filled dot, upper limits with a blue triangle.
    The black solid line shows the best fitted model. The inset shows the colour $gri$ RGB image of the galaxy with overlayed the red boxes 
    indicating the three fields used to study the SED with the CIGALE fitting code. 
 }
   \label{SED}%
   \end{figure*}

The H$\alpha$ image of NGC 4424 obtained during the VESTIGE survey clearly indicates that the activity of star formation is present only within
the inner $\sim$ 1.5 $\times$ 1.5 kpc$^2$ region, while it is totally quenched in the outer disc. Figure \ref{multifrequency} also shows that the colour
gradually increases with galactic radius (the size of the stellar disc increases with wavelength). The lack of diffuse dust in the outer regions,
shown by the 24 $\mu$m image, indicates that the radial reddening of the disc is not due to dust attenuation but it is rather produced by the aging of the stellar population.
The VLA HI data suggest that the atomic
gas in the outer disc has been removed during a ram pressure episode. The lack of gas has induced the decrease of the star formation activity.
We use the spectro-photometric data to measure the epoch when the activity has been quenched by the interaction
with the ICM. This is done by dating the last episod of star formation in the outer disc. To do that we fit the $FUV$-to-far-IR SED of the galaxy extracted in three distinct
0.5 $\times$ 0.5 arcmin$^2$ regions (2.5 $\times$ 2.5 kpc$^2$) located along the major axis of the galaxy, as depicted in Fig. \ref{SED}. These regions have been 
defined to be representative of the inner star forming region and of the outer quenched disc. 
Fluxes and uncertainties are measured within these regions using the same procedures described in Fossati et al. (2018) and Boselli et al. (2018b). 


To quantify the time elapsed since the truncation of 
the star formation activity we follow the same procedure given in Boselli et al. (2016b), i.e. we use an abruptly truncated star formation law
with as free parameters: the rotational velocity of the galaxy, the quenching age $QA$ (epoch when the quenching 
of the star formation activity started to occur) and the quenching factor $QF$ ($QF$ = 0 for unperturbed SFR, $QF$ = 1 for a totally quenched SFR). The SED of the galaxy is fitted
with CIGALE (Noll et al. 2009; Boquien et al. in prep.) using Bruzual \& Charlot (2003) stellar population models derived with a Chabrier IMF for the stellar continuum and the Draine \& Li (2007) models
for the dust emission. To represent the observed SED, we use the two GALEX $FUV$ and $NUV$ bands, the optical NGVS and VESTIGE $ugriz$
bands, a Calar Alto
near-infrared observation in the $H$ band, the four IRAC bands, the MIPS 24 $\mu$m image, the PACS images at 100 and 160 $\mu$m, and the SPIRE 250 and 350
$\mu$m bands (17 photometric bands). We exclude from the analysis the WISE bands, the MIPS bands at longer wavelengths, and the SPIRE 500 $\mu$m band for their lower
sensitivity and/or lower angular resolution to avoid any possible contamination within the selected regions. Given their size, we do not apply any aperture correction.
As in Boselli et al. (2016b) we include
the H$\alpha$ emission by measuring the number of ionising photons from the VESTIGE image. The observed H$\alpha$+[NII] flux measured within the
central region is corrected for dust attenuation and [NII] contamination using the mean values derived from the MUSE data (see Sect. 4.3). 
We also include in the fit the age-sensitive H$\beta$ absorption line
(Poggianti et al. 1998) by measuring its contribution within a pseudo-filter as in Boselli et al. (2016b). This is done using a light-weighted mean spectrum
derived from the MUSE data cube in the central and eastern regions. 

   \begin{figure*}
   \centering
   \includegraphics[width=0.8\textwidth]{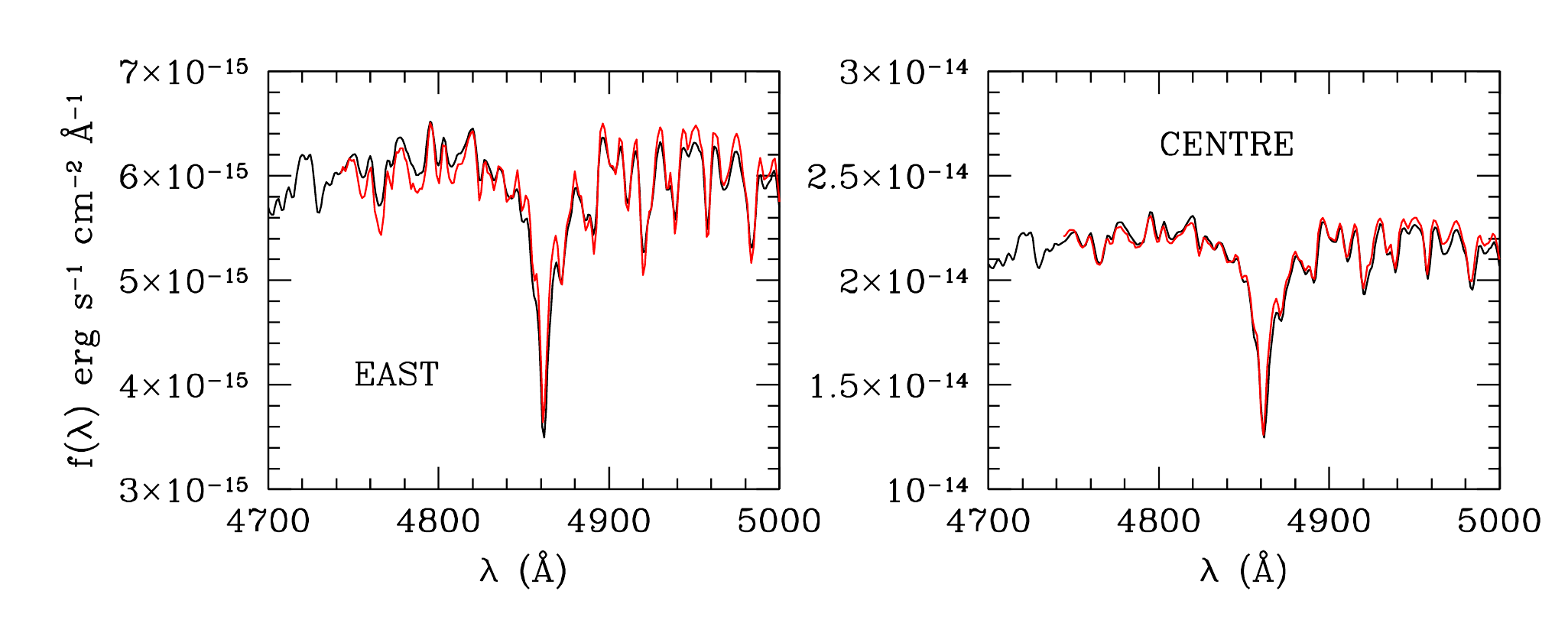}
   \caption{The best model fit obtained with CIGALE (black) is compared to the MUSE spectrum (red) in the H$\beta$ line once the emission is removed using GANDALF.
   The best model has been shifted on the Y-axis to match the observed spectrum.}
   \label{spettro}%
   \end{figure*}

The results of the fit are shown in Fig. \ref{SED} and Fig. \ref{spettro} and summarised in Table \ref{TabSED}. The quality of the fit is satisfactory in the two
outer regions, and good in the centre of the galaxy. The best fitted models perfectly match the age-sensitive H$\beta$ absorption line in the eastern and central region where MUSE data are
available.
The fitted model indicates that the activity of star formation has been totally quenched
($QF$=1) in the outer disc $\simeq$ 250-280 Myr ago. These numbers are consistent with those derived by Crowl \& Kenney (2008) using Balmer absorption indices 
extracted from IFU spectroscopy or the GALEX $FUV-NUV$
coulour index in the outer disc in the eastern direction (275 $\pm$ 75 Myr), albeit with a much better accuracy thanks to our larger number of photometric (17) and
spectroscopic (2) bands and to the more
refined SED fitting technique. To test the reliability of the output parameters of the fit, and in particular of the quenching ages and quenching factors
given above, we create as in Boselli et al. (2016b) a mock catalogue by introducing in the observed data some extra noise randomly distributed according to a Gaussian curve of
dispersion similar to the median error in each band, and re-fit the data using the SED models. The output of the fit on the mock catalogue gives again $QF$ = 1 
and $QA$ values consistent within $\simeq$ 6\%, confirming that the determination of these parameters is robust.
The fit also suggests that, in the central region, the activity of star formation has been reduced by 79\% ~ approximately 180 Myr ago. The mock analysis suggests that,
while in this region $QF$ is fairly well constrained, the uncertainty on the quenching age $QA$ is very large. Furthermore, this value should be
considered with caution given that the parametrised star formation history used to trace the evolution of NGC 4424 is not optimal for the inner
region, which has probably suffered from two competitive mechanisms, a quenching episode (during the ram pressure stripping event) and a starburst (during the
merging event) which might last 10-100 Myr (Cortijo-Ferrero et al. 2017). This very peculiar and changing star formation history is hardly reproducible using a simple SED fitting technique as the one adopted
for the outer disc. To conclude, the overall star formation activity of NGC 4424 has been significantly reduced after the gas stripping episode, 
moving the galaxy below the main sequence relation of unperturbed objects (Boselli et al. 2016b).

\begin{table*}
\caption{Results of the SED fitting}
\label{TabSED}
{
\[
\begin{tabular}{cccccc}
\hline
\noalign{\smallskip}
\hline
Region		& R.A.(J2000)	&	Dec	&QF     	& QA (Myr)		& SFR (M$_{\odot}$ yr$^{-1}$) \\      
\hline
West$^a$	& 12:27:09.001	& +9:25:19.06	&1		&	281$\pm$27 	& 0		\\
Centre		& 12:27:11.478	& +9:25:15.27	&0.79$\pm$0.04	&	179$\pm$68	& 0.13$\pm$0.02	\\
East		& 12:27:13.955	& +9:25:10.70	&1		&	251$\pm$14 	& 0		\\
\noalign{\smallskip}
\hline
\end{tabular}
\]
Note: a) fit done without the H$\beta$ pseudo filter.}
\end{table*}

\subsection{Ram pressure stripping}

The presence of a long ($\sim$ 110 kpc in projected distance) symmetric HI tail of gas is a clear evidence that the galaxy is undergoing a ram pressure stripping event 
(Chung et al. 2007, Sorgho et al. 2017). The relative orientation of the galaxy and the position of the tail on the plane of the sky, combined with 
a relatively low recessional velocity with respect to the mean velocity of the cluster ($<$$vel_{cluster A}$ - $vel_{N4424}$$>$ $\sim$ 500 km s$^{-1}$) also indicate that the stripping
event is occurring almost face-on, making it particularly efficient at removing the gas from the outer regions producing truncated gaseous discs (Quilis et al. 2000; 
Roediger \& Bruggen 2006, 2007; Tonnesen \ Bryan 2009). Given its accurate distance estimate of $\simeq$ 15.5-16 Mpc 
(Munari et al. 2013, Cantiello et al. 2018, Hatt et al. 2018), the galaxy is now in the foreground of the cluster (16.5 Mpc).
It has thus probably already crossed the southern envelopes of Virgo cluster A coming from the
south-eastern region behind the cluster and moving towards the north-west direction in front of it.
The lack of gas quenches the activity of star formation outside-in, producing truncated discs also in the young stellar populations 
(Boselli et al. 2006, Fossati et al. 2018). The SED fitting analysis presented in the previous section indicates that the quenching episode is relatively recent ($\simeq$ 250-280 Myr
ago). This timescale can be compared with the time necessary for the galaxy to travel within the cluster to form an HI tail of $\simeq$ 110 kpc. Assuming the typical velocity dispersion 
measured within Virgo cluster A ($\sigma_{cluster A}$ = 800 km s$^{-1}$; Boselli et al. 2014b), this timescale is $t_{HI}$ $\simeq$ 140 Myr, just a factor of $\sim$ 2 shorter than the typical 
quenching time. These numbers are fairly consistent considering that i) the total extension of the HI tail can be underestimated for sensitivity reasons and ii)
the cold atomic gas, once stripped and in contact with the hot intracluster medium can change phase becoming ionised gas or hot plasma via heat conduction, as indeed observed in the majority
of cluster galaxies (e.g. Boselli \& Gavazzi 2014, Boselli et al. 2016a, Yagi et al. 2017, Gavazzi et al. 2018). Within the stellar disc the gas can be shocked-ionised by the stripping process, as observed 
in the outer regions of the MUSE field (Fig. \ref{BPT}). The significant difference between the HI mass in the tail measured by Sorgho et al. (2017; $M_{tail}(HI)$ $\simeq$ 5.1 $\times$
10$^7$ M$_{\odot}$) and that of the stripped gas derived from the HI-deficiency parameter ($HI-def$ $=$ 0.98; Boselli et al. 2014c, thus 
$M_{stripped}(HI)$ $\simeq$ 2 $\times$ 10$^9$ M$_{\odot}$) supports this interpretation.

\subsection{Merging}

The morphological and kinematical properties of NGC 4424 also suggest that the galaxy has undergone a more violent gravitational perturbation 
(Kenney et al. 1996, Cortes et al. 2006, 2008, 2015). Although the presence of fine structures witnessing an old major merging event 
is not confirmed in our NGVS data, the perturbed and decoupled kinematics of both the gas and stellar components unquestionably indicate an unrelaxed object. 
In this regard, we recall the presence of a decoupled kinematics of the stellar component along the elongated
star forming structure with respect to the slowly rotating stellar disc revealed by MUSE.
Simulations suggest that unequal-mass merger remnants might result in systems with a spiral-like morphology but elliptical-like kinematics (Bournaud et al. 2004), 
characteristics similar to those observed in NGC 4424: a grand-design spiral morphology (Fig. 1)
and a low rotational velocity compared to the velocity dispersion of both the gas and the stellar components.
The fact that the gas is still turbulent and not stabilised over the disc suggests that the merging event is relatively recent ($t_{merging}$ $\lesssim$ 500 Myr).
Considering that the typical crossing time of Virgo is $\simeq$ 1.7 Gyr (Boselli \& Gavazzi 2006), this means that the merging episode
occurred once the galaxy was already within the Virgo cluster. We recall that in massive clusters such as Virgo 
($M_{Virgo}$ $\simeq$ 2-4 $\times$ 10$^{14}$ M$_{\odot}$, paper I, and references therein) 
the high velocity dispersion ($\sigma_{Virgo}$ $\simeq$ 800 km s$^{-1}$) makes merging events among intermediate mass galaxies very unlikely (Boselli \& Gavazzi 2006).

\subsection{The nature of the northern ionised gas tail}

The presence of a tail of ionised gas in the northern direction, thus in the direction opposite to the direction of the HI gas tail, 
is hard to explaine in a simple ram pressure stripping scenario. The kinematics of the gas over the plane of the disc, as mentioned in Sect. 4.3,
can be interpreted as due to i) a rotating gas ring, ii) to infalling or iii) outflowing gas. It is hard to discriminate between these different scenarios, in particular because
the peculiar morphology of the galaxy prevents us from stating which is the orientation of the stellar disc on the plane of the sky.
There are, however, a few arguments suggesting that this ionised gas tail is resulting from an outflow: \\
a) the morphology of the inner regions is similar to that observed in other starburst galaxies such as M82 or the Circinus galaxy: the H$\alpha$ emitting gas shows 
prominent filaments perpendicular to the stellar disc (Fig. \ref{Ha}), dust filamentary structures (Fig. \ref{HST}) in the inner regions, while extended tails
in H$\alpha$ (Fig. \ref{Ha}) and in the mid-IR (Fig. \ref{multifrequency}) at large scales (up to 10 kpc) 
also present in these iconic objects (Elmouttie et al. 1998a, 1998b, Shopbell \& Bland-Hawthorn 1998, Alton et al. 1999, Wilson et al. 2000, Engelbracht et al. 2006).\\
b) the velocity dispersion of the gas increases from the centre of the galaxy along the minor axis, reaching values as high as $\simeq$ 80 km s$^{-1}$ at $\simeq$ 2 kpc from the disc
(e.g. Cresci et al. 2017; Lopez-Coba et al. 2017).\\ 
c) the galaxy has two extended radio continuum lobes and a magnetic field perpendicular to the stellar disc, interpreted by Vollmer et al. (2013) as an evidence of outflow.\\
d) the flat radio continuum spectral index ($\alpha$ = -0.26; Vollmer et al. 2013) and the BPT diagram (Fig. \ref{BPT}) consistently indicate that the source of energy is a nuclear starburst.
The energy of this starburst, although now limited (0.25 M$_{\odot}$ yr$^{-1}$), could have been sufficient to overcome the gravitational forces also thanks to the perturbed gravitational potential 
well of the galaxy and to eject matter in the direction perpendicular to the disc. Following Boselli et al. (2016a) we can make a rough estimate of the total (potential and kinetic) 
energy of the outflow, $E_{out}$ $\simeq$ 1.7 $\times$ 10$^{55}$ ergs. This energy can be produced by 1.7 $\times$ 10$^4$ supernovae, or a larger number if we assume a more realistic 
1-10\% ~ energy transfer efficiency, associated to a nuclear star cluster of
stellar mass $\gtrsim$  3.7 $\times$ 10$^6$ M$_{\odot}$ and to a star formation rate $\gtrsim$ 3.7 M$_{\odot}$ yr$^{-1}$, a star formation rate comparable
to the one that the galaxy had before the quenching episode (Sect. 5.1). As remarked in Boselli et al. (2016a), however, ram pressure can produce low-density superbubble
holes and thus supply extra energy to the outflow through Kelvin-Helmholtz instabilities (Roediger \& Henseler 2005) and viscous stripping (Roediger \& Bruggen 2008).\\
e) the presence of a nuclear outflow might decouple the rotation of the stellar disc from the kinematics of the gas which becomes turbulent, as indeed observed in blue compact galaxies (Ostlin et al. 2004).\\
f) the recombination time of the gas within the tail is very low ($t_{rec}$ $\simeq$ 1.4 Myrs). It is thus unlikely that it remains ionised if it is gas orbiting along 
the minor axis or if it is infalling gas given the lack of any ionising source in these regions. In the case of an outflow, the gas can be ionised by the central starburst
or by the diffuse hot gas emitting in X-rays via heat conduction.
The BPT diagram shown in Fig. \ref{BPT}, however, indicates that at the periphery of the central star forming region detected by MUSE the gas is mainly shock-ionised probably by the ram pressure episode. \\
g) the outflow has sufficient kinetic energy to overcome the external pressure of the ICM ($p_{RP}$) if:

\begin{equation}
{p_{out} \geq p_{RP}}
\end{equation}

\noindent
where $p_{out}$ is the pressure within the outflow. A rough estimate of $p_{out}$ can be derived as:

\begin{equation}
{p_{out} = \rho_{out} vel_{out}^2 \simeq \frac{M_{out}}{V_{out}}  vel_{out}^2}
\end{equation}

\noindent
where $V_{out}$ is the volume, $M_{out}$ the mass, and $vel_{out}$ the velocity of the outflow. As described in Sect. 4.1.2, the mass of the outflow is $M_{out}$ = 6 $\times$ 10$^6$ M$_{\odot}$ 
and its volume $V$ $=$ 3.9 $\times$ 10$^{66}$ cm$^3$. We do not have any estimate of the velocity of the gas up to 10 kpc from the galaxy disc, but we can estrapolate it 
from the MUSE data (Sect. 4.3.2, Fig. \ref{velMUSE}): the gas velocity gradually increases from 0 km s$^{-1}$ up to $\frac{50 \rm{km s^{-1}}}{cos(i)}$ at $\simeq$ 2 kpc
from the galaxy disc, and might thus reach $vel_{out}$ $\simeq$ $\frac{250 \rm{km s^{-1}}}{cos(i)}$ $\simeq$ 500 km s$^{-1}$ at the edge of the observed outflow (10 kpc). 
This velocity is close to the escape velocity of the galaxy, that we estimate as in Boselli et al. (2018b) at $v_e$ $\simeq$ 700 km s$^{-1}$. 
The mean pressure exerted by the outflow on the surrounding medium should thus be $p_{out}$ $\simeq$ 7.5 $\times$ 10$^{-12}$ dyn cm$^{-2}$.
Considering the typical velocity of the galaxy within Virgo cluster A ($vel$ = 800 km s$^{-1}$, Boselli et al. 2014b), and estimating the density
of the ICM in the position of NGC 4424 (at a physical distance of 1.3 Mpc from M87) using \textit{Planck} and \textit{Suzaku} data from Simionescu et al. (2017), we obtain 
$n_{e}(r_{N4424})$ = 3 $\times$ 10$^{-5}$ cm$^{-3}$ and $p_{RP}$ $=$ 6.2 $\times$ 10$^{-13}$ dyn cm$^{-2}$ 




The accurate distance estimate (15.5-16.0 Mpc) and its relative velocity of $\simeq$ - 500 km s$^{-1}$ with respect to M87 place
the galaxy in the foreground of the cluster and suggest that it has probably crossed the periphery of Virgo coming from the south-east
and moving towards the north-west direction. If the H$\alpha$ tail is an outflow, the upper edge of the disc is the one closest to us. The H$\alpha$ tail, as the 
HI tail, is bent to the east because of the motion of the galaxy along the line-of-sight. The line-of-sight velocity of the HI gas in the tail is $vel_{tail}(HI)$ = 453 km s$^{-1}$,
slightly higher than that of the HI gas over the stellar disc ($vel_{disc}(HI)$ = 433 km s$^{-1}$; Sorgho et al. 2017), again suggesting that the gas 
has been lost while the galaxy was approaching from the cluster background, as indeed observed in other ram pressure stripped galaxies (Fumagalli et al. 2014).
Remarkably, the ionised gas in the z-plane in the south of the galaxy disc, as observed by MUSE and by the Fabry-Perot observations, has a lower line-of-sight velocity 
than the galaxy, again suggesting that the outflow overcomes the ram pressure stripping process also in this southern region.

The new result of this analysis
is that, despite a moderate nuclear star formation activity probably induced by the infall of fresh gas accreted
after a recent merging event, the galactic winds produced in the star forming regions are able to overcome the ram pressure stripping
process and drag the gas out from the disc of the galaxy from the nuclear regions even in the direction opposite to that of the motion within the cluster. This picture is fairly consistent with the prediction of
cosmological simulations and semi-analytic models in a starvation scenario (e.g. Larson et al. 1980), where the removal of the hot halo once the galaxy becomes satellite of a larger 
dark matter structure makes the feedback process more and more efficient in removing the cold gas from the galaxy disc and quenching the activity of star formation (e.g. De Lucia 2011). 
It must be noticed, however, that as in NGC 4569 (Boselli et al. 2016a) the amount of ionised gas ejected by feedback is one-to-two orders of magnitudes lower
than that removed by the ram pressure mechanism.

\section{Conclusion}

NGC 4424 is a peculiar galaxy located at $\sim$ 0.6 $R_{200}$ from the Virgo cluster centre.
New very deep NB H$\alpha$ imaging data acquired during the VESTIGE survey revealed the presence of a 
$\sim$ 10 kpc long tail (projected distance) of ionised gas perpendicular to the stellar disc. This newly discovered tail is located in the direction 
opposite to the $\sim$ 110 kpc long HI tail observed at 21 cm and generally interpreted as a clear evidence of an ongoing ram pressure stripping event.
The detailed analysis of this object based on multifrequency data covering the whole electromagnetic spectrum, from the X-rays to the radio continuum,
including high- and medium-resolution Fabry-Perot and MUSE spectroscopic data, consistently indicate that the galaxy is indeed suffering an ongoing ram pressure stripping event
able to remove the gas from the outer disc and quench the activity of star formation on timescales of $\sim$ 250-280 Myr. 
The galaxy, however, has been also perturbed by a recent merging event. This gravitational perturbation, as suggested by simulations, might have contributed to make the ram pressure stripping event more efficient
than in other cluster galaxies. The tail of ionised gas observed in the opposite direction than the HI tail results probably from a nuclear outflow 
fed by the collapse of the gas in the central regions after the merging episode. Despite a moderate nuclear star formation activity, the outflow 
produced by the starburst is able to overcome the ram pressure stripping force and contribute to the exhaustion of the gas in the inner regions.
This observational result is a first evidence that the feedback process generally invoked by cosmological simulations and semi-analytic models (e.g. De Lucia 2011) in a 
starvation scenario (e.g. Larson et al. 1980) might concur to remove gas from galaxies once they become satellites of a larger dark matter structure.
As observed in other cluster objects, however, the amount of gas expelled by feedback seems to be only a small fraction of that 
removed via ram pressure stripping, suggesting thus that feedback plays only a minor role in quenching the activity of star formation of galaxies in high density environments.


\begin{acknowledgements}

We thank the anonymous referee for constructive comments and suggestions.
We are grateful to the whole CFHT QSO team who assisted us in the preparation and in the execution of the observations and helped us with the calibration
and characterisation of the instrument, allowing us to make the best use of our
data: Todd Burdullis, Daniel Devost, Billy Mahoney, Nadine Manset, Andreea
Petric, Simon Prunet, Kanoa Withington. The authors thank the anonimous referee for the useful comments which helped improving the quality of the manuscript.
The scientific results reported in this article are based in part on data obtained from the Chandra Data Archive.
We acknowledge financial support from "Programme National de Cosmologie et Galaxies" (PNCG) funded by CNRS/INSU-IN2P3-INP, CEA and CNES, France, and from
"Projet International de Coop\'eration Scientifique" (PICS) with Canada funded by the CNRS, France.
M.B. acknowledges the support of FONDECYT regular grant 1170618.
Parts of this research were conducted by the Australian Research Council Centre of Excellence for All Sky Astrophysics in 3 Dimensions (ASTRO 3D), through project number CE170100013.
M.F. acknowledges support by the Science and Technology Facilities Council [grant number ST/P000541/1]. This project has received funding from the European Research 
Council (ERC) under the European Union's Horizon 2020 research and innovation programme (grant agreement No 757535).
SFS is grateful for the support of a CONACYT (Mexico) grant CB-285080, and funding from the PAPIIT-DGAPA-IA101217(UNAM).
M.S. acknowledges support from the NASA grant 80NSSC18K0606 and the NSF grant 1714764.
This research has made use of the NASA/IPAC Extragalactic Database (NED) 
which is operated by the Jet Propulsion Laboratory, California Institute of 
Technology, under contract with the National Aeronautics and Space Administration
and of the GOLDMine database (http://goldmine.mib.infn.it/) (Gavazzi et al. 2003b).

\end{acknowledgements}


\begin{thebibliography}{}

\bibitem[Abramson \& Kenney(2014)]{2014AJ....147...63A} Abramson, A., \& Kenney, J.~D.~P.\ 2014, \aj, 147, 63 
\bibitem[Abramson et al.(2011)]{2011AJ....141..164A} Abramson, A., Kenney, J.~D.~P., Crowl, H.~H., et al.\ 2011, \aj, 141, 164 
\bibitem[Abramson et al.(2016)]{2016AJ....152...32A} Abramson, A., Kenney, J., Crowl, H., \& Tal, T.\ 2016, \aj, 152, 32 
\bibitem[Alton et al.(1999)]{1999A&A...343...51A} Alton, P.~B., Davies, J.~I., \& Bianchi, S.\ 1999, \aap, 343, 51 
\bibitem[Baldwin et al.(1981)]{1981PASP...93....5B} Baldwin, J.~A., Phillips, M.~M., \& Terlevich, R.\ 1981, \pasp, 93, 5 
\bibitem[Bendo et al.(2012)]{2012MNRAS.423..197B} Bendo, G.~J., Galliano, F., \& Madden, S.~C.\ 2012, \mnras, 423, 197 
\bibitem[Binggeli et al.(1985)]{1985AJ.....90.1681B} Binggeli, B., Sandage, A., \& Tammann, G.~A.\ 1985, \aj, 90, 1681 
\bibitem[Blakeslee et al.(2009)]{2009ApJ...694..556B} Blakeslee, J.~P., Jord{\'a}n, A., Mei, S., et al.\ 2009, \apj, 694, 556 
\bibitem{2006PASP..118..517B} Boselli, A., \& Gavazzi, G.\ 2006, \pasp, 118, 517 
\bibitem[Boselli \& Gavazzi(2014)]{2014A&ARv..22...74B} Boselli, A., \& Gavazzi, G.\ 2014, \aapr, 22, 74 
\bibitem[Boselli et al.(2000)]{2000A&AS..142...73B} Boselli, A., Gavazzi, G., Franzetti, P., Pierini, D., \& Scodeggio, M.\ 2000, \aaps, 142, 73 
\bibitem[Boselli et al.(2005)]{2005ApJ...623L..13B} Boselli, A., Boissier, S., Cortese, L., et al.\ 2005, \apjl, 623, L13 
\bibitem{2006ApJ...651..811B} Boselli, A., Boissier, S., Cortese, L., et al.\ 2006, \apj, 651, 811 
\bibitem[Boselli et al.(2009)]{2009ApJ...706.1527B} Boselli, A., Boissier, S., Cortese, L., et al.\ 2009, \apj, 706, 1527 
\bibitem[Boselli et al.(2010)]{2010PASP..122..261B} Boselli, A., Eales, S., Cortese, L., et al.\ 2010, \pasp, 122, 261 
\bibitem{2011A&A...528A.107B} Boselli, A., Boissier, S., Heinis, S., et al.\ 2011, \aap, 528, A107 
\bibitem[Boselli et al.(2014)]{2014A&A...564A..67B} Boselli, A., Cortese, L., Boquien, M., et al.\ 2014a, \aap, 564, A67 
\bibitem[Boselli et al.(2014)]{2014A&A...570A..69B} Boselli, A., Voyer, E., Boissier, S., et al.\ 2014b, \aap, 570, AA69 
\bibitem[Boselli et al.(2014)]{2014A&A...564A..65B} Boselli, A., Cortese, L., \& Boquien, M.\ 2014c, \aap, 564, A65 
\bibitem[Boselli et al.(2015)]{2015A&A...579A.102B} Boselli, A., Fossati, M., Gavazzi, G., et al.\ 2015, \aap, 579, A102 
\bibitem[Boselli et al.(2016)]{2016A&A...587A..68B} Boselli, A., Cuillandre, J.~C., Fossati, M., et al.\ 2016a, \aap, 587, A68 
\bibitem[Boselli et al.(2016)]{2016A&A...596A..11B} Boselli, A., Roehlly, Y., Fossati, M., et al.\ 2016b, \aap, 596, A11 
\bibitem[Boselli et al.(2018)]{2018A&A...614A..56B} Boselli, A., Fossati, M., Ferrarese, L., et al.\ 2018, \aap, 614, A56 
\bibitem[]{} Boselli, Fossati, Cuillandre, et al. 2018b, arXiv:1803.04177
\bibitem[Bournaud et al.(2004)]{2004A&A...418L..27B} Bournaud, F., Combes, F., \& Jog, C.~J.\ 2004, \aap, 418, L27 
\bibitem[Bruzual \& Charlot(2003)]{2003MNRAS.344.1000B} Bruzual, G., \& Charlot, S.\ 2003, \mnras, 344, 1000 
\bibitem[Byrd \& Valtonen(1990)]{1990ApJ...350...89B} Byrd, G., \& Valtonen, M.\ 1990, \apj, 350, 89 
\bibitem[Cantiello et al.(2018)]{2018ApJ...856..126C} Cantiello, M., Blakeslee, J.~P., Ferrarese, L., et al.\ 2018, \apj, 856, 126 
\bibitem[Cappellari \& Emsellem(2004)]{2004PASP..116..138C} Cappellari, M., \& Emsellem, E.\ 2004, \pasp, 116, 138 
\bibitem[Chabrier(2003)]{2003PASP..115..763C} Chabrier, G.\ 2003, \pasp, 115, 763 
\bibitem[Chung et al.(2007)]{2007ApJ...659L.115C} Chung, A., van Gorkom, J.~H., Kenney, J.~D.~P., \& Vollmer, B.\ 2007, \apjl, 659, L115 
\bibitem[Chung et al.(2009)]{2009AJ....138.1741C} Chung, A., van Gorkom, J.~H., Kenney, J.~D.~P., Crowl, H., \& Vollmer, B.\ 2009, \aj, 138, 1741 
\bibitem[Chy{\.z}y et al.(2006)]{2006A&A...447..465C} Chy{\.z}y, K.~T., Soida, M., Bomans, D.~J., et al.\ 2006, \aap, 447, 465 
\bibitem[Ciesla et al.(2012)]{2012A&A...543A.161C} Ciesla, L., Boselli, A., Smith, M.~W.~L., et al.\ 2012, \aap, 543, A161 
\bibitem[Ciesla et al.(2014)]{2014A&A...565A.128C} Ciesla, L., Boquien, M., Boselli, A., et al.\ 2014, \aap, 565, A128 
\bibitem[Coccato et al.(2005)]{2005A&A...440..107C} Coccato, L., Corsini, E.~M., Pizzella, A., \& Bertola, F.\ 2005, \aap, 440, 107 
\bibitem[Consolandi et al.(2017)]{2017A&A...606A..83C} Consolandi, G., Gavazzi, G., Fossati, M., et al.\ 2017, \aap, 606, A83 
\bibitem[Cort{\'e}s et al.(2006)]{2006AJ....131..747C} Cort{\'e}s, J.~R., Kenney, J.~D.~P., \& Hardy, E.\ 2006, \aj, 131, 747 
\bibitem[Cort{\'e}s et al.(2008)]{2008ApJ...683...78C} Cort{\'e}s, J.~R., Kenney, J.~D.~P., \& Hardy, E.\ 2008, \apj, 683, 78-93 
\bibitem[Cort{\'e}s et al.(2015)]{2015ApJS..216....9C} Cort{\'e}s, J.~R., Kenney, J.~D.~P., \& Hardy, E.\ 2015, \apjs, 216, 9 
\bibitem[Cortese et al.(2006)]{2006A&A...453..847C} Cortese, L., Gavazzi, G., Boselli, A., et al.\ 2006, \aap, 453, 847 
\bibitem[Cortese et al.(2010)]{2010A&A...518L..49C} Cortese, L., Davies, J.~I., Pohlen, M., et al.\ 2010, \aap, 518, L49 
\bibitem[Cortese et al.(2012a)]{2012A&A...540A..52C} Cortese, L., Ciesla, L., Boselli, A., et al.\ 2012a, \aap, 540, A52 
\bibitem[Cortese et al.(2012b)]{2012A&A...544A.101C} Cortese, L., Boissier, S., Boselli, A., et al.\ 2012b, \aap, 544, A101 
\bibitem[Cortese et al.(2014)]{2014MNRAS.440..942C} Cortese, L., Fritz, J., Bianchi, S., et al.\ 2014, \mnras, 440, 942 
\bibitem[Cortijo-Ferrero et al.(2017)]{2017A&A...607A..70C} Cortijo-Ferrero, C., Gonz{\'a}lez Delgado, R.~M., P{\'e}rez, E., et al.\ 2017, \aap, 607, A70 
\bibitem[Cowie \& Songaila(1977)]{1977Natur.266..501C} Cowie, L.~L., \& Songaila, A.\ 1977, \nat, 266, 501 
\bibitem[Cresci et al.(2017)]{2017A&A...604A.101C} Cresci, G., Vanzi, L., Telles, E., et al.\ 2017, \aap, 604, A101 
\bibitem[Crowl \& Kenney(2008)]{2008AJ....136.1623C} Crowl, H.~H., \& Kenney, J.~D.~P.\ 2008, \aj, 136, 1623 
\bibitem[Curti et al.(2017)]{2017MNRAS.465.1384C} Curti, M., Cresci, G., Mannucci, F., et al.\ 2017, \mnras, 465, 1384 
\bibitem[Daigle et al.(2006)]{2006MNRAS.368.1016D} Daigle, O., Carignan, C., Hernandez, O., Chemin, L., \& Amram, P.\ 2006, \mnras, 368, 1016 
\bibitem[Davies et al.(2010)]{2010A&A...518L..48D} Davies, J.~I., Baes, M., Bendo, G.~J., et al.\ 2010, \aap, 518, L48 
\bibitem[De Lucia(2011)]{2011ASSP...27..203D} De Lucia, G.\ 2011, Astrophysics and Space Science Proceedings, 27, 203 
\bibitem[Dressler(1980)]{1980ApJ...236..351D} Dressler, A.\ 1980, \apj, 236, 351 
\bibitem[Dressler et al.(1997)]{1997ApJ...490..577D} Dressler, A., Oemler, A., Jr., Couch, W.~J., et al.\ 1997, \apj, 490, 577 
\bibitem[Duc et al.(2011)]{2011MNRAS.417..863D} Duc, P.-A., Cuillandre, J.-C., Serra, P., et al.\ 2011, \mnras, 417, 863 
\bibitem[Elmouttie et al.(1998)]{1998MNRAS.297...49E} Elmouttie, M., Koribalski, B., Gordon, S., et al.\ 1998a, \mnras, 297, 49 
\bibitem[Elmouttie et al.(1998)]{1998MNRAS.297.1202E} Elmouttie, M., Haynes, R.~F., Jones, K.~L., Sadler, E.~M., \& Ehle, M.\ 1998b, \mnras, 297, 1202 
\bibitem[Engelbracht et al.(2006)]{2006ApJ...642L.127E} Engelbracht, C.~W., Kundurthy, P., Gordon, K.~D., et al.\ 2006, \apjl, 642, L127 
\bibitem[Epinat et al.(2008)]{2008MNRAS.388..500E} Epinat, B., Amram, P., Marcelin, M., et al.\ 2008, \mnras, 388, 500 
\bibitem[Ferrarese et al.(2012)]{2012ApJS..200....4F} Ferrarese, L., C{\^o}t{\'e}, P., Cuillandre, J.-C., et al.\ 2012, \apjs, 200, 4 
\bibitem[Fossati et al.(2012)]{2012A&A...544A.128F} Fossati, M., Gavazzi, G., Boselli, A., \& Fumagalli, M.\ 2012, \aap, 544, A128 
\bibitem[Fossati et al.(2016)]{2016MNRAS.455.2028F} Fossati, M., Fumagalli, M., Boselli, A., et al.\ 2016, \mnras, 455, 2028 
\bibitem[Fossati et al.(2018)]{2018A&A...614A..57F} Fossati, M., Mendel, J.~T., Boselli, A., et al.\ 2018, \aap, 614, A57 
\bibitem[Fritz et al.(2017)]{2017ApJ...848..132F} Fritz, J., Moretti, A., Gullieuszik, M., et al.\ 2017, \apj, 848, 132 
\bibitem[Fumagalli et al.(2009)]{2009ApJ...697.1811F} Fumagalli, M., Krumholz, M.~R., Prochaska, J.~X., Gavazzi, G., \& Boselli, A.\ 2009, \apj, 697, 1811 
\bibitem[Fumagalli et al.(2011)]{2011A&A...528A..46F} Fumagalli, M., Gavazzi, G., Scaramella, R., \& Franzetti, P.\ 2011, \aap, 528, A46 
\bibitem[Fumagalli et al.(2014)]{2014MNRAS.445.4335F} Fumagalli, M., Fossati, M., Hau, G.~K.~T., et al.\ 2014, \mnras, 445, 4335 
\bibitem[Gach et al.(2002)]{2002PASP..114.1043G} Gach, J.-L., Hernandez, O., Boulesteix, J., et al.\ 2002, \pasp, 114, 1043 
\bibitem[Galbany et al.(2016)]{2016MNRAS.455.4087G} Galbany, L., Anderson, J.~P., Rosales-Ortega, F.~F., et al.\ 2016, \mnras, 455, 4087 
\bibitem[Galbany et al.(2018)]{2018MNRAS.tmp.1394G} Galbany, L., Collett, T.~E., M{\'e}ndez-Abreu, J., et al.\ 2018, \mnras,  
\bibitem[Garrido et al.(2005)]{2005MNRAS.362..127G} Garrido, O., Marcelin, M., Amram, P., et al.\ 2005, \mnras, 362, 127 
\bibitem[Gavazzi et al.(1995)]{1995A&A...304..325G} Gavazzi, G., Contursi, A., Carrasco, L., et al.\ 1995, \aap, 304, 325 
\bibitem[Gavazzi et al.(1998)]{1998AJ....115.1745G} Gavazzi, G., Catinella, B., Carrasco, L., Boselli, A., \& Contursi, A.\ 1998, \aj, 115, 1745 
\bibitem[Gavazzi et al.(1999)]{1999MNRAS.304..595G} Gavazzi, G., Boselli, A., Scodeggio, M., Pierini, D., \& Belsole, E.\ 1999, \mnras, 304, 595 
\bibitem[Gavazzi et al.(2001)]{2001ApJ...563L..23G} Gavazzi, G., Boselli, A., Mayer, L., et al.\ 2001a, \apjl, 563, L23 
\bibitem[Gavazzi et al.(2001)]{2001A&A...377..745G} Gavazzi, G., Marcelin, M., Boselli, A., et al.\ 2001b, \aap, 377, 745 
\bibitem[Gavazzi et al.(2002)]{2002ApJ...576..135G} Gavazzi, G., Bonfanti, C., Sanvito, G., Boselli, A., \& Scodeggio, M.\ 2002, \apj, 576, 135 
\bibitem[Gavazzi et al.(2003)]{2003ApJ...597..210G} Gavazzi, G., Cortese, L., Boselli, A., et al.\ 2003a, \apj, 597, 210 
\bibitem[Gavazzi et al.(2003)]{2003A&A...400..451G} Gavazzi, G., Boselli, A., Donati, A., Franzetti, P., \& Scodeggio, M.\ 2003b, \aap, 400, 451 
\bibitem[Gavazzi et al.(2005)]{2005A&A...429..439G} Gavazzi, G., Boselli, A., van Driel, W., \& O'Neil, K.\ 2005, \aap, 429, 439 
\bibitem[Gavazzi et al.(2006)]{2006A&A...449..929G} Gavazzi, G., O'Neil, K., Boselli, A., \& van Driel, W.\ 2006a, \aap, 449, 929 
\bibitem[Gavazzi et al.(2006)]{2006A&A...446..839G} Gavazzi, G., Boselli, A., Cortese, L., et al.\ 2006b, \aap, 446, 839 
\bibitem[Gavazzi et al.(2013)]{2013A&A...553A..89G} Gavazzi, G., Fumagalli, M., Fossati, M., et al.\ 2013, \aap, 553, A89 
\bibitem[]{} Gavazzi, Consolandi, Gutierrez, Boselli, Yoshida, 2018, in prep.
\bibitem[]{}Ge C., Morandi, A., Sun, M., et al., 2018: arXiv:1803.05007
\bibitem[Gunn \& Gott(1972)]{1972ApJ...176....1G} Gunn, J.~E., \& Gott, J.~R., III 1972, \apj, 176, 1 
\bibitem[Gwyn(2008)]{2008PASP..120..212G} Gwyn, S.~D.~J.\ 2008, \pasp, 120, 212
\bibitem[Hatt et al.(2018)]{2018ApJ...861..104H} Hatt, D., Freedman, W.~L., Madore, B.~F., et al.\ 2018, \apj, 861, 104 
\bibitem[Haynes et al.(1984)]{1984ARA&A..22..445H} Haynes, M.~P., Giovanelli, R., \& Chincarini, G.~L.\ 1984, \araa, 22, 445 
\bibitem[Hughes et al.(2013)]{2013A&A...550A.115H} Hughes, T.~M., Cortese, L., Boselli, A., Gavazzi, G., \& Davies, J.~I.\ 2013, \aap, 550, A115 
\bibitem[J{\'a}chym et al.(2013)]{2013A&A...556A..99J} J{\'a}chym, P., Kenney, J.~D.~P., R{\v z}ui{\v c}ka, A., et al.\ 2013, \aap, 556, A99 
\bibitem[J{\'a}chym et al.(2014)]{2014ApJ...792...11J} J{\'a}chym, P., Combes, F., Cortese, L., Sun, M., \& Kenney, J.~D.~P.\ 2014, \apj, 792, 11 
\bibitem[Kandrashoff et al.(2012)]{2012CBET.3111....1K} Kandrashoff, M., Cenko, S.~B., Li, W., et al.\ 2012, Central Bureau Electronic Telegrams, 3111, 
\bibitem[Kauffmann et al.(2003)]{2003MNRAS.346.1055K} Kauffmann, G., Heckman, T.~M., Tremonti, C., et al.\ 2003, \mnras, 346, 1055 
\bibitem[Kenney \& Koopmann(1999)]{1999AJ....117..181K} Kenney, J.~D.~P., \& Koopmann, R.~A.\ 1999, \aj, 117, 181 
\bibitem[Kenney \& Yale(2002)]{2002ApJ...567..865K} Kenney, J.~D.~P., \& Yale, E.~E.\ 2002, \apj, 567, 865 
\bibitem[Kenney et al.(1996)]{1996AJ....111..152K} Kenney, J.~D.~P., Koopmann, R.~A., Rubin, V.~C., \& Young, J.~S.\ 1996, \aj, 111, 152 
\bibitem[Kenney et al.(2004)]{2004AJ....127.3361K} Kenney, J.~D.~P., van Gorkom, J.~H., \& Vollmer, B.\ 2004, \aj, 127, 3361 
\bibitem[Kenney et al.(2008)]{2008ApJ...687L..69K} Kenney, J.~D.~P., Tal, T., Crowl, H.~H., Feldmeier, J., \& Jacoby, G.~H.\ 2008, \apjl, 687, L69 
\bibitem[Kenney et al.(2014)]{2014ApJ...780..119K} Kenney, J.~D.~P., Geha, M., J{\'a}chym, P., et al.\ 2014, \apj, 780, 119 
\bibitem[Kenney et al.(2015)]{2015AJ....150...59K} Kenney, J.~D.~P., Abramson, A., \& Bravo-Alfaro, H.\ 2015, \aj, 150, 59 
\bibitem[Kennicutt(1983)]{1983AJ.....88..483K} Kennicutt, R.~C., Jr.\ 1983, \aj, 88, 483 
\bibitem[Kennicutt(1998)]{1998ARA&A..36..189K} Kennicutt, R.~C., Jr.\ 1998, \araa, 36, 189 
\bibitem[Kewley et al.(2001)]{2001ApJ...556..121K} Kewley, L.~J., Dopita, M.~A., Sutherland, R.~S., Heisler, C.~A., \& Trevena, J.\ 2001, \apj, 556, 121 
\bibitem[Kr{\"u}hler et al.(2017)]{2017A&A...602A..85K} Kr{\"u}hler, T., Kuncarayakti, H., Schady, P., et al.\ 2017, \aap, 602, A85 
\bibitem[Larson et al.(1980)]{1980ApJ...237..692L} Larson, R.~B., Tinsley, B.~M., \& Caldwell, C.~N.\ 1980, \apj, 237, 692 
\bibitem[L{\'o}pez-Cob{\'a} et al.(2017)]{2017MNRAS.467.4951L} L{\'o}pez-Cob{\'a}, C., S{\'a}nchez, S.~F., Moiseev, A.~V., et al.\ 2017, \mnras, 467, 4951 
\bibitem[Marconi \& Hunt(2003)]{2003ApJ...589L..21M} Marconi, A., \& Hunt, L.~K.\ 2003, \apjl, 589, L21 
\bibitem[Mei et al.(2007)]{2007ApJ...655..144M} Mei, S., Blakeslee, J.~P., C{\^o}t{\'e}, P., et al.\ 2007, \apj, 655, 144 
\bibitem[Mendes de Oliveira et al.(2017)]{2017MNRAS.469.3424M} Mendes de Oliveira, C., Amram, P., Quint, B.~C., et al.\ 2017, \mnras, 469, 3424 
\bibitem[Merritt(1983)]{1983ApJ...264...24M} Merritt, D.\ 1983, \apj, 264, 24 
\bibitem[Mihos et al.(2015)]{2015ApJ...809L..21M} Mihos, J.~C., Durrell, P.~R., Ferrarese, L., et al.\ 2015, \apjl, 809, L21 
\bibitem[Moore et al.(1998)]{1998ApJ...495..139M} Moore, B., Lake, G., \& Katz, N.\ 1998, \apj, 495, 139 
\bibitem[Munari et al.(2013)]{2013NewA...20...30M} Munari, U., Henden, A., Belligoli, R., et al.\ 2013, \na, 20, 30 
\bibitem[Noll et al.(2009)]{2009A&A...507.1793N} Noll, S., Burgarella, D., Giovannoli, E., et al.\ 2009, \aap, 507, 1793
\bibitem[Nulsen(1982)]{1982MNRAS.198.1007N} Nulsen, P.~E.~J.\ 1982, \mnras, 198, 1007 
\bibitem[Osterbrock \& Ferland(2006)]{2006agna.book.....O} Osterbrock, D.~E., \& Ferland, G.~J.\ 2006, Astrophysics of gaseous nebulae and active galactic nuclei, 2nd.~ed.~by D.E.~Osterbrock and G.J.~Ferland.~Sausalito, CA: University Science Books, 2006,  
\bibitem[{\"O}stlin et al.(2004)]{2004A&A...419L..43O} {\"O}stlin, G., Cumming, R.~J., Amram, P., et al.\ 2004, \aap, 419, L43 
\bibitem[Paudel et al.(2017)]{2017ApJ...834...66P} Paudel, S., Smith, R., Duc, P.-A., et al.\ 2017, \apj, 834, 66 
\bibitem[Poggianti et al.(2017)]{2017ApJ...844...48P} Poggianti, B.~M., Moretti, A., Gullieuszik, M., et al.\ 2017, \apj, 844, 48 
\bibitem[Proxauf et al.(2014)]{2014A&A...561A..10P} Proxauf, B., {\"O}ttl, S., \& Kimeswenger, S.\ 2014, \aap, 561, A10 
\bibitem[Quilis et al.(2000)]{2000Sci...288.1617Q} Quilis, V., Moore, B., \& Bower, R.\ 2000, Science, 288, 1617 
\bibitem[Rich et al.(2011)]{2011ApJ...734...87R} Rich, J.~A., Kewley, L.~J., \& Dopita, M.~A.\ 2011, \apj, 734, 87 
\bibitem[Roediger \& Hensler(2005)]{2005A&A...433..875R} Roediger, E., \& Hensler, G.\ 2005, \aap, 433, 875 
\bibitem[Roediger \& Br{\"u}ggen(2006)]{2006MNRAS.369..567R} Roediger, E., \& Br{\"u}ggen, M.\ 2006, \mnras, 369, 567 
\bibitem[Roediger \& Br{\"u}ggen(2007)]{2007MNRAS.380.1399R} Roediger, E., \& Br{\"u}ggen, M.\ 2007, \mnras, 380, 1399 
\bibitem[Roediger \& Br{\"u}ggen(2008)]{2008MNRAS.388L..89R} Roediger, E., \& Br{\"u}ggen, M.\ 2008, \mnras, 388, L89 
\bibitem[Rubin et al.(1999)]{1999AJ....118..236R} Rubin, V.~C., Waterman, A.~H., \& Kenney, J.~D.~P.\ 1999, \aj, 118, 236 
\bibitem[Sarazin(1986)]{1986RvMP...58....1S} Sarazin, C.~L.\ 1986, Reviews of Modern Physics, 58, 1 
\bibitem[Sarzi et al.(2006)]{2006MNRAS.366.1151S} Sarzi, M., Falc{\'o}n-Barroso, J., Davies, R.~L., et al.\ 2006, \mnras, 366, 1151 
\bibitem[Schindler et al.(1999)]{1999A&A...343..420S} Schindler, S., Binggeli, B., B{\"o}hringer, H.\ 1999, \aap, 343, 420 
\bibitem[Shopbell \& Bland-Hawthorn(1998)]{1998ApJ...493..129S} Shopbell, P.~L., \& Bland-Hawthorn, J.\ 1998, \apj, 493, 129 
\bibitem[Silverman et al.(2012)]{2012ApJ...756L...7S} Silverman, J.~M., Ganeshalingam, M., Cenko, S.~B., et al.\ 2012, \apjl, 756, L7 
\bibitem[Simionescu et al.(2017)]{2017MNRAS.469.1476S} Simionescu, A., Werner, N., Mantz, A., Allen, S.~W., \& Urban, O.\ 2017, \mnras, 469, 1476 
\bibitem[Solanes et al.(2001)]{2001ApJ...548...97S} Solanes, J.~M., Manrique, A., Garc{\'{\i}}a-G{\'o}mez, C., et al.\ 2001, \apj, 548, 97 
\bibitem[Sorgho et al.(2017)]{2017MNRAS.464..530S} Sorgho, A., Hess, K., Carignan, C., \& Oosterloo, T.~A.\ 2017, \mnras, 464, 530 
\bibitem[Soto et al.(2016)]{2016MNRAS.458.3210S} Soto, K.~T., Lilly, S.~J., Bacon, R., Richard, J., \& Conseil, S.\ 2016, \mnras, 458, 3210 
\bibitem[Sun et al.(2006)]{2006ApJ...637L..81S} Sun, M., Jones, C., Forman, W., et al.\ 2006, \apjl, 637, L81 
\bibitem[Sun et al.(2007)]{2007ApJ...671..190S} Sun, M., Donahue, M., \& Voit, G.~M.\ 2007, \apj, 671, 190 
\bibitem[Sun et al.(2010)]{2010ApJ...708..946S} Sun, M., Donahue, M., Roediger, E., et al.\ 2010, \apj, 708, 946 
\bibitem[Tonnesen \& Bryan(2009)]{2009ApJ...694..789T} Tonnesen, S., \& Bryan, G.~L.\ 2009, \apj, 694, 789 
\bibitem[Tonnesen et al.(2011)]{2011ApJ...731...98T} Tonnesen, S., Bryan, G.~L., \& Chen, R.\ 2011, \apj, 731, 98 
\bibitem[Vazdekis et al.(2010)]{2010MNRAS.404.1639V} Vazdekis, A., S{\'a}nchez-Bl{\'a}zquez, P., Falc{\'o}n-Barroso, J., et al.\ 2010, \mnras, 404, 1639 
\bibitem[Vollmer(2003)]{2003A&A...398..525V} Vollmer, B.\ 2003, \aap, 398, 525 
\bibitem[Vollmer et al.(1999)]{1999A&A...349..411V} Vollmer, B., Cayatte, V., Boselli, A., Balkowski, C., \& Duschl, W.~J.\ 1999, \aap, 349, 411 
\bibitem[Vollmer et al.(2000)]{2000A&A...364..532V} Vollmer, B., Marcelin, M., Amram, P., et al.\ 2000, \aap, 364, 532 
\bibitem[Vollmer et al.(2001)]{2001ApJ...561..708V} Vollmer, B., Cayatte, V., Balkowski, C., \& Duschl, W.~J.\ 2001a, \apj, 561, 708 
\bibitem[Vollmer et al.(2001)]{2001A&A...374..824V} Vollmer, B., Braine, J., Balkowski, C., Cayatte, V., \& Duschl, W.~J.\ 2001b, \aap, 374, 824 
\bibitem[Vollmer et al.(2004)]{2004AJ....127.3375V} Vollmer, B., Beck, R., Kenney, J.~D.~P., \& van Gorkom, J.~H.\ 2004a, \aj, 127, 3375 
\bibitem[Vollmer et al.(2004)]{2004A&A...419...35V} Vollmer, B., Balkowski, C., Cayatte, V., van Driel, W., \& Huchtmeier, W.\ 2004b, \aap, 419, 35 
\bibitem[Vollmer et al.(2005)]{2005A&A...439..921V} Vollmer, B., Huchtmeier, W., \& van Driel, W.\ 2005a, \aap, 439, 921 
\bibitem[Vollmer et al.(2005)]{2005A&A...441..473V} Vollmer, B., Braine, J., Combes, F., \& Sofue, Y.\ 2005b, \aap, 441, 473 
\bibitem[Vollmer et al.(2006)]{2006A&A...453..883V} Vollmer, B., Soida, M., Otmianowska-Mazur, K., et al.\ 2006, \aap, 453, 883 
\bibitem[Vollmer et al.(2008)]{2008A&A...483...89V} Vollmer, B., Soida, M., Chung, A., et al.\ 2008a, \aap, 483, 89 
\bibitem[Vollmer et al.(2008)]{2008A&A...491..455V} Vollmer, B., Braine, J., Pappalardo, C., \& Hily-Blant, P.\ 2008b, \aap, 491, 455 
\bibitem[Vollmer et al.(2009)]{2009A&A...496..669V} Vollmer, B., Soida, M., Chung, A., et al.\ 2009, \aap, 496, 669 
\bibitem[Vollmer et al.(2012)]{2012A&A...537A.143V} Vollmer, B., Soida, M., Braine, J., et al.\ 2012, \aap, 537, A143 
\bibitem[Vollmer et al.(2013)]{2013A&A...553A.116V} Vollmer, B., Soida, M., Beck, R., et al.\ 2013, \aap, 553, A116 
\bibitem[Yagi et al.(2007)]{2007ApJ...660.1209Y} Yagi, M., Komiyama, Y., Yoshida, M., et al.\ 2007, \apj, 660, 1209 
\bibitem[Yagi et al.(2010)]{2010AJ....140.1814Y} Yagi, M., Yoshida, M., Komiyama, Y., et al.\ 2010, \aj, 140, 1814 
\bibitem[Yagi et al.(2017)]{2017ApJ...839...65Y} Yagi, M., Yoshida, M., Gavazzi, G., et al.\ 2017, \apj, 839, 65 
\bibitem[Yoshida et al.(2002)]{2002ApJ...567..118Y} Yoshida, M., Yagi, M., Okamura, S., et al.\ 2002, \apj, 567, 118 
\bibitem[Weilbacher et al.(2014)]{2014ASPC..485..451W} Weilbacher, P.~M., Streicher, O., Urrutia, T., et al.\ 2014, Astronomical Data Analysis Software and Systems XXIII, 485, 451 
\bibitem[Whitmore et al.(1993)]{1993ApJ...407..489W} Whitmore, B.~C., Gilmore, D.~M., \& Jones, C.\ 1993, \apj, 407, 489 
\bibitem[Wilson et al.(2000)]{2000AJ....120.1325W} Wilson, A.~S., Shopbell, P.~L., Simpson, C., et al.\ 2000, \aj, 120, 1325 
\bibitem[Wright et al.(2010)]{2010AJ....140.1868W} Wright, E.~L., Eisenhardt, P.~R.~M., Mainzer, A.~K., et al.\ 2010, \aj, 140, 1868-1881 
\bibitem[Zhang et al.(2013)]{2013ApJ...777..122Z} Zhang, B., Sun, M., Ji, L., et al.\ 2013, \apj, 777, 122



\end{thebibliography}
\end{document}